\documentclass[fleqn,usenatbib]{mnras}

\usepackage{newtxtext,newtxmath}
\usepackage{ulem}
\usepackage[T1]{fontenc}

\DeclareRobustCommand{\VAN}[3]{#2}
\let\VANthebibliography\thebibliography
\def\thebibliography{\DeclareRobustCommand{\VAN}[3]{##3}\VANthebibliography}

\usepackage{graphicx}	% Including figure files
\usepackage{bbm}    % The use of indicator symbol 
\usepackage{enumitem}
\usepackage{subcaption}

\usepackage{xcolor}
\usepackage{xspace}

\newcommand{\gaia}{\textit{Gaia}\xspace}
\newcommand{\code}[1]{\textsc{#1}\xspace} %MNRAS style is small caps
\newcommand{\acknowledgeSoftware}[1]{\code{#1} \citep{#1}\xspace}

\newcommand{\Mv}{M_{\rm V}}
\newcommand{\FeH}{\rm \left[Fe/H\right]}

\title[RR Lyrae in ZTF DR3]{Identifying RR Lyrae in the ZTF DR3 dataset}

\author[K.-W. Huang et al.]{
Kuan-Wei Huang$^{1}$\thanks{E-mail: kuanweih@andrew.cmu.edu} and Sergey E. Koposov$^{2,3,1}$
\\
$^{1}$McWilliams Center for Cosmology, Dept. of Physics, Carnegie Mellon University, Pittsburgh, PA, 15213, USA\\
$^{2}$Institute for Astronomy, University of Edinburgh, Royal Observatory, Blackford Hill, Edinburgh EH9 3HJ, UK\\
$^{3}$Institute of Astronomy, University of Cambridge, Madingley Road, Cambridge CB3 0HA, UK
}

\date{Accepted XXX. Received YYY; in original form ZZZ}

\pubyear{2015}

\begin{document}
\label{firstpage}
\pagerange{\pageref{firstpage}--\pageref{lastpage}}
\maketitle

\begin{abstract}
We present a RR Lyrae (RRL) catalogue based on the combination of the third data release of the Zwicky Transient Facility (ZTF DR3) and \textit{Gaia} EDR3. 
We use a multi-step classification pipeline relying on the Fourier decomposition fitting to the multi-band ZTF light curves and random forest classification. 
The resulting catalogue contains 71,755 RRLs with period and light curve parameter measurements and has completeness of 0.92 and purity of 0.92 with respect to the SOS \textit{Gaia} DR2 RRLs. 
The catalogue covers the Northern sky with declination $\geq -28^\circ$, its completeness is $\gtrsim 0.8$ for heliocentric distance $\leq 80$~kpc, and the most distant RRL at 132~kpc. 
Compared with several other RRL catalogues covering the Northern sky, our catalogue has more RRLs around the Galactic halo and is more complete at low Galactic latitude areas. 
Analysing the spatial distribution of RRL in the catalogue reveals the previously known major over-densities of the Galactic halo, such as the Virgo over-density and the Hercules-Aquila Cloud, with some evidence of an association between the two.  
We also analyse the Oosterhoff fraction differences throughout the halo, comparing it with the density distribution, finding increasing Oosterhoff I fraction at the elliptical radii between 16 and 32 kpc and some evidence of different Oosterhoff fractions across various halo substructures.
\end{abstract}


%%%%%%%%%%%%%%%%%%%%%%%%%%%%%%%%%%%%%%%%%%%%%%%%%%

%%%%%%%%%%%%%%%%% BODY OF PAPER %%%%%%%%%%%%%%%%%%

\section{Introduction}
\label{sec:intro}

RR Lyrae (RRL) stars are pulsating variables with periodic light curves of a period ranging from 0.2 to 0.9 days \citep{1995CAS....27.....S}, found primarily in the horizontal branches of old stellar systems (age $>10$ Gyr). 
These old, metal-poor ($\FeH < -0.5$), bright ($\Mv = 0.59$ at $\FeH = -1.5$; \citet{Cacciari2003}) variable stars follow a well-understood period-luminosity-metallicity (PLZ) relation \citep[e.g.][]{C_ceres_2008, 2012MSAIS..19..138M}. 
This relation makes RRLs excellent distance indicators for old, low-metallicity stellar populations in the outer halo of the Milky Way \citep[e.g.][]{Catelan_2004, 2004AJ....127.1158V, C_ceres_2008, 2010ApJ...708..717S, 2014PASP..126..616S, 2015ApJ...798L..12F}. 
Besides, RRLs are sufficiently luminous to be detected at large distances so that they can be the tracer of the halo substructures with a good spatial resolution \citep[e.g.][]{Vivas_2006, 2010ApJ...708..717S, Sesar_2014, Baker_2015, 2015MNRAS.446.2251T, 10.1093/mnras/stz2609}. 
Proposed by \citet{Sesar_2014} \citep[see also][]{Baker_2015}, the fact that almost every Milky Way dwarf satellite galaxy has at least one RRL star opens up a gate of locating the Milky Way dwarf satellites even for the ones that are very faint by using distant RRL stars, for example, Antlia 2 \citep{2019MNRAS.488.2743T}.

Being beneficial to many Galactic studies, there have been several RRL catalogues classified from existing surveys over the years, e.g. SDSS Stripe 82 \citep{2010ApJ...708..717S}, CRTS \citep{2014ApJS..213....9D}, PS1 \citep{2017AJ....153..204S}, nTransits:2+ \gaia DR2 \citep{2018A&A...618A..30H}, SOS \gaia DR2 \citep{2019A&A...622A..60C}, ZTF DR2 \citep{2020ApJS..249...18C}, and DES Y6 \citep{2021ApJ...911..109S}. 
The quality of the catalogues has progressed from being either deep with limited sky coverage (e.g. the SDSS Stripe 82 catalogue) or wide-coverage but not as deep (e.g. the CRTS catalogue) to having decent depth and wide sky coverage at the same time (e.g. the PS1 catalogue), pushing the Galactic studies furthermore. 
However, large-coverage and deep surveys usually suffer from significant incompleteness and contamination due to the low number of epochs in the light curves. 
This motivates us to identify a RRL catalogue from the ZTF survey thanks to its uniformly high number of observation epochs of light curves across the Northern sky while having decent depth. 
Another challenge of the catalogues is to cover the Galactic plane; the PS1, \gaia DR2, and ZTF DR2 catalogues do cover this area though the \gaia catalogue suffers the completeness issue here. 
The PS1 data suffer the issues of sparse temporal coverage, cadence, and asynchronous multi-band observations where they overcame them by the multi-stage classification in \citet{2016ApJ...817...73H}. 
Compared to the ZTF DR2 catalogue \citep{2020ApJS..249...18C}, the more recent data release used in this work provides more observation epochs which is beneficial for the light curve fitting to achieve more accurate period measurement. 
Also in this work for the period determination, we used all the bands simultaneously during the light curve fitting stage.

In this paper, we utilize the joint set of the \gaia early third data release \citep[\gaia EDR3;][]{2020arXiv201201533G} and the third data release of the Zwicky Transient Facility \citep[ZTF DR3;][]{2019PASP..131a8003M} to classify RRL stars in the Northern sky. 
Thanks to the high angular resolution of \gaia and the fast cadence of ZTF observations, the sources in the joint set thus have high spatial resolution and multi-band light curves with large observation epochs. 
Assisted with the Specific Objects Study (SOS) \gaia DR2 RRL catalogue as the label, we process the dataset following the pipeline we come up with, which includes data labelling, feature building, and classifier training, to obtain the predicted RRL catalogue. 
In Section~\ref{sec:method_datasets}, we describe the datasets above in more detail. 
In Section~\ref{sec:method_pipeline}, we explain the pipeline step by step. 
In Section~\ref{sec:result}, we demonstrate the classification results and present the predicted RRL catalogue. 
In Section~\ref{sec:conclusion}, we conclude the paper.

\section{Datasets}
\label{sec:method_datasets}
% --------------------------------------------------

% Fig ----------------------------------------------
\begin{figure}
    \includegraphics[width=\columnwidth]{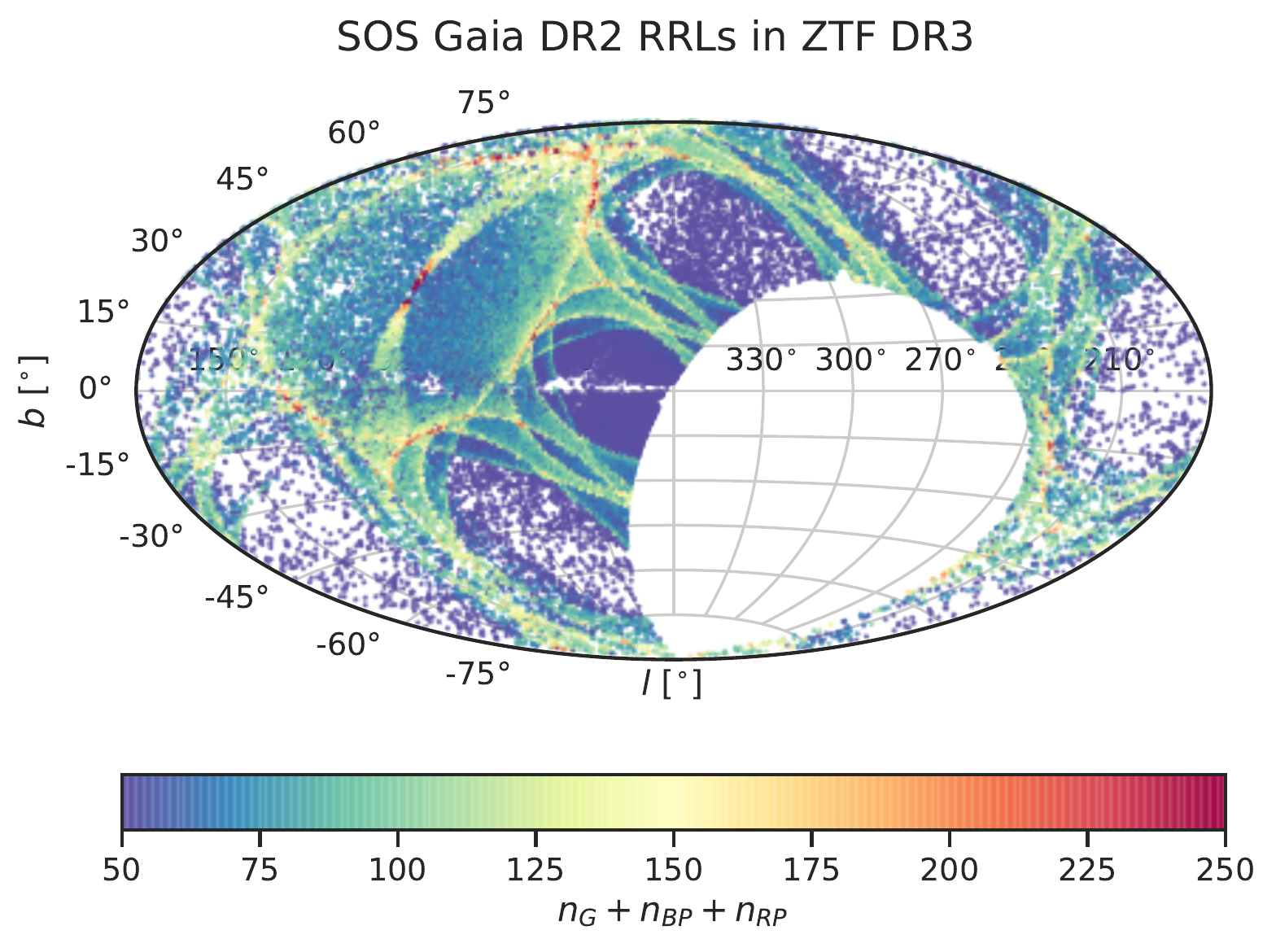}
    \caption{The spatial distribution of 48,365 SOS \gaia DR2 RRLs with detected ZTF DR3 light curves by the closest separation within one arcsec on the sky, colour-coded by the total number of \gaia epochs.} 
    \label{fig:gaia_rrls_n_epochs_pos}
\end{figure}
% Fig ----------------------------------------------

To identify RRLs in the Northern sky, we utilize three datasets in this work: ZTF DR3, \gaia EDR3, and the SOS \gaia DR2 RRL catalogue. 
The joint set of ZTF DR3 and \gaia EDR3 is the main dataset and the SOS \gaia DR2 RRL catalogue serves as the label for training models.

\textbf{ZTF DR3} \citep{2019PASP..131a8002B}: 
As a time-domain survey using the 48-inch Schmidt telescope equipped with a 47 squared degree camera at Palomar Observatory, ZTF started scanning the entire Northern sky in March 2018, covering the area of $\sim 3 \pi$ steradians. 
In the Northern sky of declination $> -31^\circ$, ZTF has conducted two surveys: the Galactic Plane Survey with a one-day cadence of all visible fields at $|b| < 7^\circ$ and the Northern Sky Survey with a three-day cadence at all fields with centres at $|b| > 7^\circ$. 
Released in June 2020, ZTF DR3 contains the data collected during the first 21.4 months of the survey and has approximately 2.5 billion light curves constructed from the single-exposure extractions, with limiting magnitudes at about $g = 20.8$, $r = 20.6$, and $i = 19.9$, and the angular resolution of about one arcsec.

\textbf{\gaia EDR3} \citep{2020arXiv201201533G}: 
The space-based astrometric mission \gaia was launched by the European Space Agency in 2013 and started the whole-sky survey in 2014 \citep{2016A&A...595A...1G}. 
Released in December 2020, \gaia EDR3 contains the data collected during the first 34 months of the mission and has approximately 1.8 billion sources with 1.5 billion parallaxes and proper motions, down to the magnitude limit of $\rm G = 20.7$. 
The angular separation limit, below which two sources are considered duplicates, has been lowered to 180 mas in EDR3, while it was 400 mas in DR2.

\textbf{The SOS \gaia DR2 RRL catalogue}: 
Using the Specific Objects Study (SOS) pipeline, \cite{2019A&A...622A..60C} presented 140,784 RRL stars in \gaia DR2 using the \gaia multi-band time-series photometry of all-sky candidate variables. 
We note that there are two RRL catalogues from Gaia DR2, the SOS catalogue and the nTransits:2+ catalogue \citep{2018A&A...618A..30H}, which is expected to be of lower quality due to a significantly smaller number of epochs per source.

To start the data preparation, we first create the joint dataset of ZTF DR3 and \gaia EDR3 by cross-matching the closest sources from the two surveys with an angular separation smaller than one arcsec. 
The resulting dataset contains 675,640,523 sources in the Northern sky down to the magnitude of about 20.5. 
The sources in the dataset thus are clearly identified but not mismatched single sources because \gaia has a higher angular resolution than ZTF. 
Each source in the joint set thus not only has the astrometric and photometric measurements from \gaia but also has the light curves in the $gri$ bands from ZTF which in particular are essential for the classification pipeline explained in the following paragraphs. 
We note that we lose about 800 million sources from the original 1,471,263,267 sources in the ZTF DR3 dataset by this cross-match mainly because ZTF is slightly deeper than \gaia in some regions, despite the similar limiting magnitudes of the two surveys.
However, the majority of the missing objects are very faint with magnitudes $> 21$ and have extremely large photometric errors.

Besides \gaia EDR3 and ZTF DR3, we use the SOS \gaia DR2 RRLs as the label for the binary classification task; we label each source in the joint dataset as true if it is classified as a RRL in the SOS \gaia DR2 RRL catalogue and as false otherwise. 
Amongst the 140,784 RRLs in the SOS \gaia DR2 RRL catalogue, 48,365 RRLs have ZTF light curves when cross-matched by the closest separation within one arcsec.  
In Figure~\ref{fig:gaia_rrls_n_epochs_pos}, we show the distribution of these 48,365 \gaia RRLs in the Galactic coordinate colour-coded by the total number of \gaia epochs, where $n_G$, $n_{BP}$, and $n_{RP}$ are \texttt{num\_clean\_epochs\_g}, \texttt{num\_clean\_epochs\_bp}, and \texttt{num\_clean\_epochs\_rp} respectively. 
Figure~\ref{fig:gaia_rrls_n_epochs_pos} illustrates the incompleteness issue that the SOS \gaia DR2 RRL catalogue suffers in the low-epoch areas due to the scanning trajectory of \gaia, which we will take into account during the classification pipeline.

% --------------------------------------------------
\section{The classification pipeline}
\label{sec:method_pipeline}
% --------------------------------------------------

With the dataset of 600 million sources in the joint set of \gaia EDR3 and ZTF DR3 and the label of the SOS \gaia DR2 RRLs, we then proceed to the supervised classification of RR Lyrae candidates through the multi-step process summarized below and described in detail in later sections.

\textbf{The initial variability selection}: 
    To make the period fitting process computationally feasible, in Section~\ref{subsec:method_initial_variability_selection}, we first reduce the size of the dataset to 155,095,514 sources by applying an initial variability selection based on the residuals of constant flux fits to the ZTF light curves.

\textbf{The broad selection of RRL candidates}:
     Since the computational cost of the full Fourier period fitting for 155 million sources is still prohibitive, in Section~\ref{subsec:method_broad_selection_RRLs}, we perform a further filtering step by doing a discretised single sinusoidal fit to characterize the periodic variability of the sources. Together with the results from the previous step, we further rule out the unlikely variable sources using a random forest classifier and end up with 3,041,677 sources.

\textbf{The final classification of RRLs}: 
    In Section~\ref{subsec:method_final_classification}, we build features for the dataset of 3 million sources using the parameters obtained by fitting truncated Fourier Series to each light curve in multiple bands. Then we train another random forest classifier to predict the probability of a source being a RRL and generate a catalogue of 71,755 RRLs.

Since we employ the ZTF light curves for every step, we here lay out the data we use before diving into the detail of the classification process. 
For each band $k = g, r, i$ in ZTF, $n_k$ is the number of ZTF detection with $\texttt{catflags} < 32768$, which flags bad or generally unusable observation epochs \citep{2019PASP..131a8003M}. 
For the $i$-th detection for $i \in \{1, 2, ..., n_k\}$, $t_{k,i}$ is the observed time \texttt{mjd\_k}, $m_{k,i}$ is the observed magnitude \texttt{mag\_k}, and $\sigma_{k,i}$ is the uncertainty of the observed magnitude \texttt{magerr\_k}.

% --------------------------------------------------
\subsection{The initial variability selection}
\label{subsec:method_initial_variability_selection}
% --------------------------------------------------

% Fig ----------------------------------------------
\begin{figure}
    \includegraphics[width=\columnwidth]{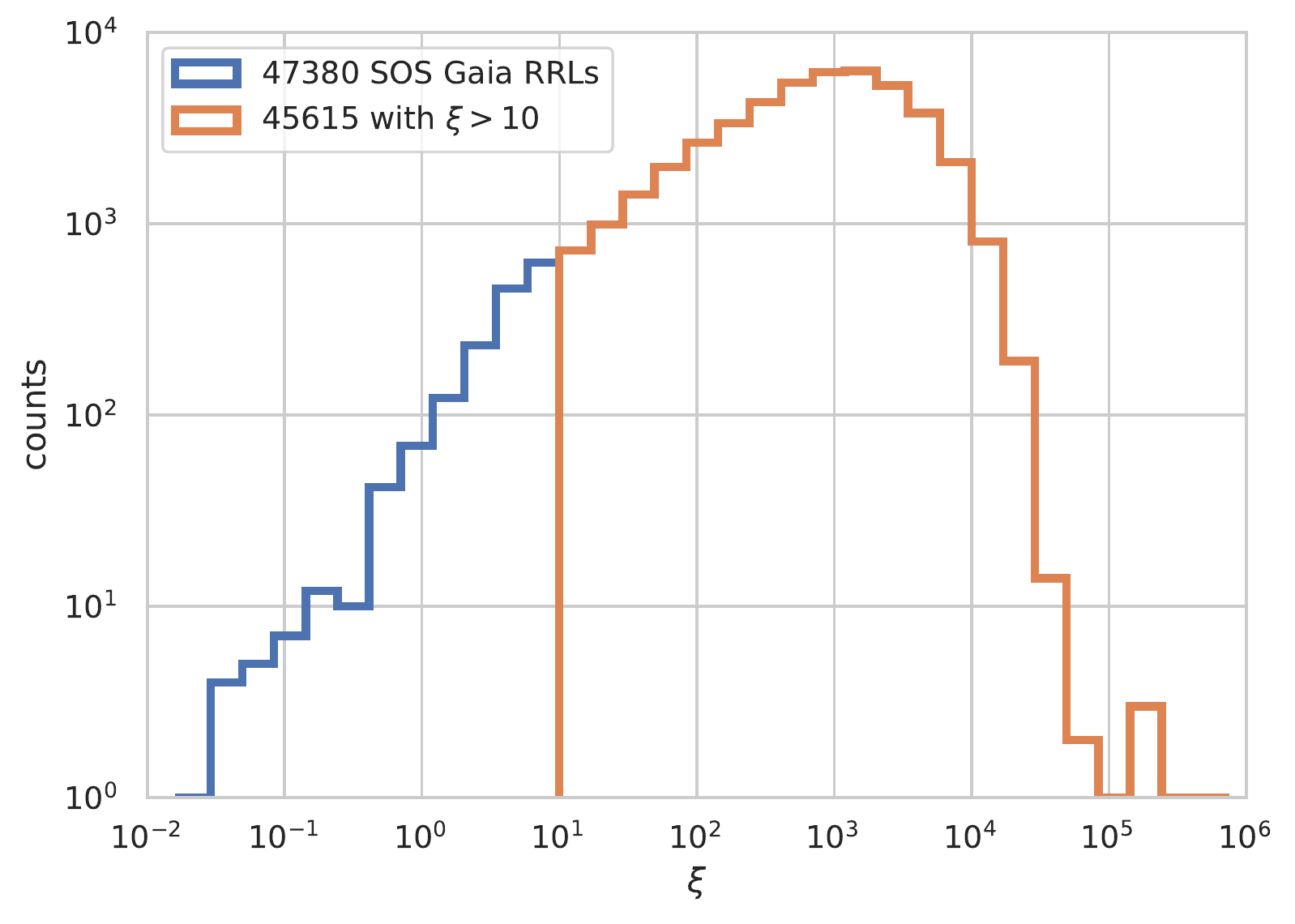}
    \caption{The distribution of SOS \gaia RRLs in terms of $\xi$ defined in Equation~\ref{eq:xi_variability}. The blue and orange histograms are before and after the selection of Equation~\ref{eq:ini_var_selection} respectively.} 
    \label{ini_var_cut_all_ztf3_gaia_rrls}
\end{figure}
% Fig ----------------------------------------------

We start to process the 600 million sources in the joint set of \gaia EDR3 and ZTF DR3 by two selections to make the size of the dataset feasible for variable light curve fittings in the following steps. 
The first selection is 
\begin{equation}
\label{eq:ini_num_good_selection}
    n_k \geq 10
\end{equation}
for any ZTF band $k = g, r, i$.
The reason is to keep the sources with at least 10 light curve data points in any given band such that the single sinusoidal fitting and the truncated Fourier fitting in the following steps are reasonable. 
After the selection of Equation~\ref{eq:ini_num_good_selection}, 47,380 out of the total 48,365 SOS \gaia RRLs in ZTF DR3 survive.

The second selection is based on the variability inferred by the residuals of constant light curve fits. 
The constant light curve model for band $k$ is defined as
\begin{equation}
\label{eq:const_model}
    m_{{\rm C}_k} (t) = C_k.
\end{equation} 
The estimator of the parameter $C_k$ is  the mean of the observed light curve data points; $C_k = \frac{1}{n_k} \sum_{i = 1}^{n_k} m_{k,i}$. 
For each light curve, we evaluate the sum of squared residuals as
\begin{equation}
\label{eq:chi_k_const}
    \chi_{{\rm C}_k}^2 = \sum_{i = 1}^{n_k} 
    \left( \frac{m_{k,i} - m_{{\rm C}_k} \left( t_{k,i} \right)}{\sigma_{k,i}} \right)^2.
\end{equation} 
Using the $g$ and $r$ band statistics, we characterize the significance of variability as a scalar quantity
\begin{equation}
\label{eq:xi_variability}
    \xi = \frac{ \chi_{{\rm C}_g}^2 + \chi_{{\rm C}_r}^2 + \nu}{\sqrt{2 \nu}}
\end{equation}
where $\nu = n_g + n_r - 2$ is the degrees of freedom, similar to Equation~1 in \citet{2016ApJ...817...73H}. 
We exclude the $i$ band because $\sim 96\%$ of the ZTF sources have $< 10$ epochs in their $i$-band light curves.
The blue histogram in Figure~\ref{ini_var_cut_all_ztf3_gaia_rrls} shows the distribution of the 47,380 SOS \gaia RRLs in terms of $\xi$. 
To keep as many SOS \gaia RRLs as possible while shrinking the size of the overall dataset as small as possible, we decide to have the cut of
\begin{equation}
\label{eq:ini_var_selection}
    \xi > 10
\end{equation}
as the second selection. 
As shown in the orange histogram in Figure~\ref{ini_var_cut_all_ztf3_gaia_rrls}, this selection keeps 45,615 from the 47,380 SOS \gaia RRLs.

After the selections of Equation~\ref{eq:ini_num_good_selection} and Equation~\ref{eq:ini_var_selection}, 155,095,514 out of the 600 million sources in the joint set of \gaia EDR3 and ZTF DR3 survive, entering the next step in the following section. 
The completeness of the SOS \gaia RRLs after the two selections of Equation~\ref{eq:ini_num_good_selection} and Equation~\ref{eq:ini_var_selection} is 0.94.

% --------------------------------------------------
\subsection{The broad selection of RRL candidates}
\label{subsec:method_broad_selection_RRLs}
% --------------------------------------------------

Because it is still too computationally expensive to perform higher-order Fourier fitting of all 155 million sources selected in the previous step, we need an extra step to further select a smaller subset of sources.
Utilizing two simple and computationally feasible models of the multiple-band ZTF light curves described in Section~\ref{subsubsec:method_features_for_rfc1}, we obtain features to characterize the periodicity and variability of the sources and train the random forest classifier I to broadly select the possible RRL candidates in Section~\ref{subsubsec:method_rfc1}.

% --------------------------------------------------
\subsubsection{Constant and single sinusoidal light curve fitting}
\label{subsubsec:method_features_for_rfc1}
% --------------------------------------------------

% Tab ----------------------------------------------
\begin{table}
\caption{The features we use to train the random forest classifier I. The total ZTF epoch $n_{\rm tot} = n_g + n_r + n_i$. $\bar{k}$ and $\tilde{k}$ are the mean and median of the $k$-band magnitude with $k=g$ and $r$. $Q_{j} (k)$ is the $j$th quartile of the $k$-band magnitude with $k=g$ and $r$.} 
\label{tab:rfc1_features}
\begin{tabular}{lll}
symbol                               & explanation                                   & range                      \\
\hline
$\log_{10} n_{\rm tot}$              & log of total ZTF epochs                       &                            \\
$\left( \bar{g} - \bar{r} \right)_0$                  & $\bar{g} - \bar{r} - E(B - V)$                &                            \\
$\left( \tilde{g} - \tilde{r} \right)_0$              & $\tilde{g} - \tilde{r} - E(B - V)$            &                            \\
$\rho_{gr}$                          & correlation of $g$ and $r$ light curves       & $\left[ -1, 1 \right]$     \\
$\rho_{gg}$                          & auto-correlation of $g$ light curves          & $\left[ -1, 1 \right]$     \\
$\rho_{rr}$                          & auto-correlation of $r$ light curves          & $\left[ -1, 1 \right]$     \\
$Q_{12} (g)$                         & $Q_{1} (g) - Q_{2} (g)$                       &                            \\
$Q_{12} (r)$                         & $Q_{1} (r) - Q_{2} (r)$                       &                            \\
$Q_{32} (g)$                         & $Q_{3} (g) - Q_{2} (g)$                       &                            \\
$Q_{32} (r)$                         & $Q_{3} (r) - Q_{2} (r)$                       &                            \\
$\delta \chi^2_{\rm S, C}$           & normalized delta chi-square in Equation~\ref{eq:dchisq_sin_const} &                            \\
$P_{\rm sin}$                        & best fitting period from single sinusoidal fit                      & $\left[ 0.1, 30 \right]$   \\
\hline
\end{tabular}
\end{table}
% Tab ----------------------------------------------

The first of the two simple and computationally feasible models is the constant light curve fit mentioned in the previous section. 
The other model is a single discretized sinusoidal light curve formulated as
\begin{equation}
    \label{eq:model_single_sin}
    m_{{\rm S}_{k,i}} = A_k \cos^* \left( \frac{2\pi}{P} t_{k,i} + \phi_k \right) + B_k
\end{equation}
where $\cos^*$ is the discretized cosine and the parameters $A_k$ and $B_k$ are the amplitudes, $\phi_k$ is the phase, and $P$ is the period. 
For each band $k$, the sum of squared residuals for the single sinusoidal model is defined as 
\begin{equation}
\label{eq:chi_k_sin}
    \chi_{{\rm S}_k}^2 = \sum_{i = 1}^{n_k} 
    \left( \frac{m_{k,i} - m_{{\rm S}_{k,i}}}{\sigma_{k,i}} \right)^2.
\end{equation}

Fitting the period ranging between 0.1 and 30 days for the light curve in each band with more than 10 ZTF detections with $\texttt{catflags} < 32768$, we have the best fits with the residual sums of squares in the multiple bands for each source for each trial period. Then we pick the fit with the best period $P_{\rm S}$ that minimizes the total residual sum of squares $\chi^2_{\rm S}$ in the multiple bands as the best fit of the single discretized sinusoidal light curve. 
This fitting process for the 155 million sources took about 300k CPU hours to complete (one month on machines of 420 cores of Intel Haswell E5-2695 v3 CPUs).

From the fits of the two models, we select a set of features summarized in Table~\ref{tab:rfc1_features} for the broad selection in the following step.
The selected features are the total number of epochs, the de-reddened colour index $g - r$ based on the mean and the median observed light curves, the difference of the magnitudes between quartiles, the correlations, the best period of the single sinusoidal fit, and the difference of the residual sum of squares $\delta \chi^2_{\rm S, C}$ defined as 
\begin{equation}
\label{eq:dchisq_sin_const}
    \delta \chi^2_{\rm S, C} = \frac{\chi^2_{\rm C} - \chi^2_{\rm S}}{\sqrt{2\chi^2_{\rm C}}},
\end{equation}
where the total residual sums of squares $\chi_{\rm C}^2 = \chi_{{\rm C}_g}^2 + \chi_{{\rm C}_r}^2 + \chi_{{\rm C}_i}^2$ and $\chi_{\rm S}^2 = \chi_{{\rm S}_g}^2 + \chi_{{\rm S}_r}^2 + \chi_{{\rm S}_i}^2$ according to Equation~\ref{eq:chi_k_const} and Equation~\ref{eq:chi_k_sin} respectively, the term of $\sqrt{2\chi^2_{\rm C}}$ is the approximate uncertainty from the variance of the chi-square distribution. 
Ideally, given a number of epochs, a source with a higher $\delta \chi^2_{\rm S, C}$ is more periodically variable than a source with a lower $\delta \chi^2_{\rm S, C}$.

% --------------------------------------------------
\subsubsection{Random forest classification I}
\label{subsubsec:method_rfc1}
% --------------------------------------------------

% Fig ----------------------------------------------
\begin{figure}
    \includegraphics[width=\columnwidth]{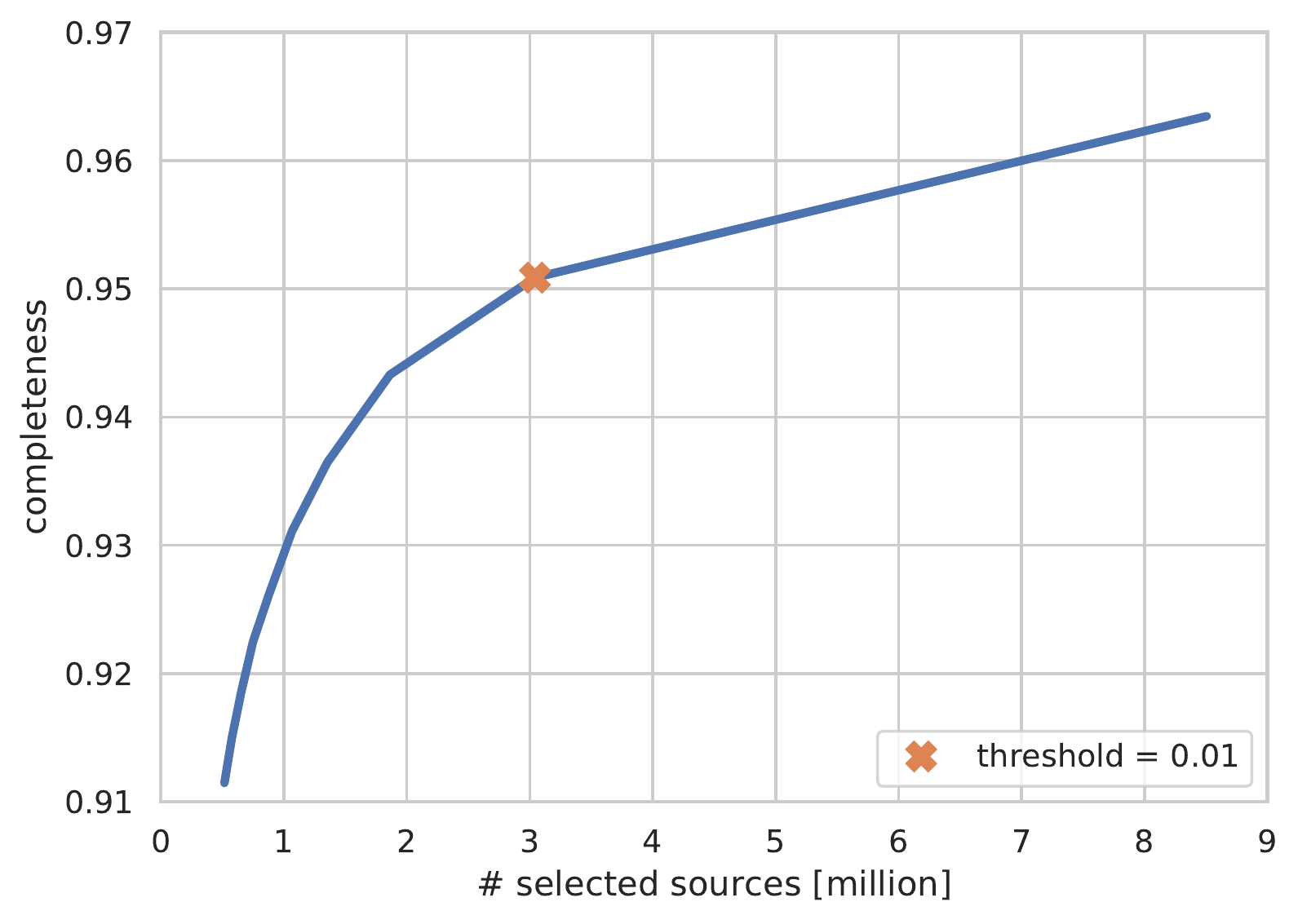}
    \caption{The relation between the completeness and the number of selected sources according to different probability thresholds ranging between 0 and 1. The orange mark shows the threshold of 0.01 which is the one we use for the random forest classification I in our pipeline.} 
    \label{fig:sfps_roc_nselect}
\end{figure}
% Fig ----------------------------------------------

With the features listed in Table~\ref{tab:rfc1_features} and the label from the SOS \gaia DR2 RRLs, we train a 10-fold cross-validation random forest classifier on the 155 million sources to identify periodic variable sources that are likely to be RRLs by predicting the probability of a source being a possible RRL candidate. 
Utilizing the random forest classifier in \code{Scikit-Learn} \citep{scikit-learn}, we employ the default parameters from the module but customize the objective function to be the cross-entropy function and the weights to be adjusted inversely proportional to class frequencies in the input data.
For details of random forests and the module, we refer readers to \citet{Statistics01randomforests} and \citet{scikit-learn}. 
The 10-fold cross-validation is done by randomly shuffling the 155 million entries and partitioning them into 10 subsets. 
For each subset, we train a classifier using the other nine subsets as the training set and use the classifier to compute the predicted probability for the subset. 
Repeating this for all 10 subsets, we accomplish the cross-validation prediction for all the 155 million sources.

Based on the cross-validation prediction, we show the completeness versus the number of selected sources with different probability thresholds between 0 and 0.1 in Figure~\ref{fig:sfps_roc_nselect}.  
Limited by our computational resources, we can only afford to fit at most roughly 3 million sources with higher-order Fourier Series in the next step, so we decide to use the probability threshold of 0.01 for the selection.
With the probability larger than 0.01, there are 3,041,677 selected sources, whose completeness is 0.95 and purity is 0.014. 
This dataset of 3 million sources then enters the final step of the pipeline described in the following sections.

% --------------------------------------------------
\subsection{The final RRL classification step}
\label{subsec:method_final_classification}
% --------------------------------------------------

Using the 3 million sources selected previously as the dataset, we are ready to process the last step in the pipeline to identify RRLs. 
We first fit each ZTF light curve with the third order of Fourier Series to find the best period and select a set of features that characterizes the shape of light curves in Section~\ref{subsubsec:method_feature_classification}. 
With the selected feature set, we train the random forest classifier II to predict the probability of each source being a RRL in Section~\ref{subsubsec:method_rfc2}.

% --------------------------------------------------
\subsubsection{Fourier Series fitting}
\label{subsubsec:method_feature_classification}
% --------------------------------------------------

% Tab ----------------------------------------------
\begin{table}
\caption{The features of the training set we use for the random forest classifier. Note that $k$ denotes $g$ or $r$ bands.}
\label{tab:features_rfc2}
\begin{tabular}{lll}
symbol                               & explanation                                         & range                      \\
\hline
$P_{\rm best}$                       & best fitting period                             & $\left[ 0.1, 1 \right]$    \\
$\left( g - r \right)_0$                              & $A_{g, 0} - A_{r, 0} - E(B - V)$                    &                            \\
$\ln A_{k, 1}$                       & log of the first Fourier amplitude                  &                            \\
$\ln A_{k, 2}$                       & log of the second Fourier amplitude                 &                            \\
$\ln A_{k, 3}$                       & log of the third Fourier amplitude                  &                            \\
$\phi_{k, 21}$                       & the second relative phase                           & $\left[ -\pi, \pi \right]$ \\
$\phi_{k, 31}$                       & the third relative phase                            & $\left[ -\pi, \pi \right]$ \\
$\delta \chi^2_{\rm F, C}$           & normalized delta chi square in Equation~\ref{eq:dchisq_fourier_const}                    &                            \\
\hline
\end{tabular}
\end{table}
% Tab ----------------------------------------------

We model each ZTF light curve in band $k$ using the third order of the Fourier Series as 
\begin{equation}
\label{eq:model_fourier_fit_ori}
    m_{{\rm F}_k} (t) = A_{k, 0} + \sum_{j=1}^{3} A_{k, j} \cos ( j \omega t + \phi_{k, j} ) 
\end{equation} 
with the parameters of the angular frequency $\omega = \frac{2 \pi}{P}$, the period $P$, the Fourier amplitudes $A_{k, 0}, A_{k, j}$ and phases $\phi_{k, j}$ for $j = \{1, 2, 3\}$. 
We note that for the objects with a large number of light curve points, the accurate description of the light curve might require more high-order Fourier terms than three. 
To fit a light curve using the model if there is more than 10 detection with $\texttt{catflags} < 32768$ for the light curve, we use a uniform grid in $\frac{1}{P}$ with $10^5$ points of the period between 0.1 and 1 days.
Given a period, we fit each light curve using the model with the lowest residual sum of squares computed as 
\begin{equation}
\label{eq:chi_k_fourier}
    \chi_{{\rm F}_k}^2 = \sum_{i = 1}^{n_k} 
    \left( \frac{m_{k,i} - m_{{\rm F}_k} \left( t_{k,i} \right)}{\sigma_{k,i}} \right)^2. 
\end{equation} 

For each trial period we perform fits to data in every band and then sum their resulting chi-squares as $\chi_{{\rm F}}^2 = \chi_{{\rm F}_g}^2 + \chi_{{\rm F}_r}^2 + \chi_{{\rm F}_i}^2$ to be the indicator for determining the best fitting result, that is, the fit with the best period $P_{\rm best}$ that minimizes $\chi_{{\rm F}}^2$. 
We note that practically we fit light curves  using the model (Eq.~\ref{eq:model_fourier_fit_ori}) for each period by doing linear regression with respect to the 1, [$\sin \left( j \omega t \right)$, $\cos \left( j \omega t \right)$] for $j = \{1, 2, 3\}$, which can be done with one single matrix operation. 
This Fourier fitting process for the 3 million sources took about 600k CPU hours to complete (two months on machines of 420 cores of Intel Haswell E5-2695 v3 CPUs).

With the fitted parameters $\left( P_{\rm best}, \  A_{k, 0} , \  A_{k, j} , \  \phi_{k, j} \right)$ for $j = \{1, 2, 3\}$, to choose features for the classifier, we aim to use the parameters that characterize the shape of light curves because of the unique shape of RRL light curves. 
The terms of the zeroth amplitude $A_{k, 0}$ and the first phase $\phi_{k, 1}$ are essentially the mean magnitude and the phase shift respectively for the light curve so they contribute no meaningful information about the shape of light curves. 
Thus we exclude them. 
Because $\phi_{k, 1}$ does affect the other phase terms, we rewrite Equation~\ref{eq:model_fourier_fit_ori} in the form of 
\begin{equation}
\begin{split}
    m_k (t) 
        &= A_{k, 0} + A_{k, 1} \cos ( \omega \tau_k) \\
        &\quad + A_{k, 2} \cos ( 2 \omega \tau_k + \phi_{k, 21} ) + A_{k, 3} \cos ( 3 \omega \tau_k + \phi_{k, 31} ) 
\end{split}
\end{equation} 
to take care of the time shift caused by $\phi_{k, 1}$, where $\tau_k = t + \frac{\phi_{k, 1}}{\omega}$, $\phi_{k, 21} = \phi_{k, 2} - 2 \phi_{k, 1}$, and $\phi_{k, 31} = \phi_{k, 3} - 3 \phi_{k, 1}$. 
Unlike $\phi_{k, 1}$, these relative phases $\phi_{k, 21}$ and $\phi_{k, 31}$ do characterize the shape of light curves so we include them in the feature set. 
It is worth noting that there is a correlation between metallicity and $\phi_{k, 31}$ \citep{1993ApJ...410..526S, 1996A&A...312..111J, 2004AJ....128..858S, 2010ApJ...708..717S}.

Besides the shape of light curves, the difference in the goodness of the Fourier fit and that of the constant light curve fit is essential to the classification because it indicates the goodness of the two competing models. 
Similar to Equation~\ref{eq:dchisq_sin_const} in Section~\ref{subsubsec:method_features_for_rfc1}, we define the normalized delta chi-square as
\begin{equation}
\label{eq:dchisq_fourier_const}
    \delta \chi^2_{\rm F, C} = \frac{\chi^2_{\rm C} - \chi^2_{\rm F}}{\sqrt{2\chi^2_{\rm C}}}
\end{equation}
and include it in the feature set, where $\chi^2_{\rm F}$ and $\chi^2_{\rm C}$ are the residual sums of squares of the best Fourier fit and that of the constant fit. 
For example, for the light curve of a true RRL, the Fourier light curve tends to fit it better than the constant light curve does, resulting in low $\chi^2_{\rm F}$, high $\chi^2_{\rm C}$, and thus a large value of $\delta \chi^2_{\rm F, C}$.

To sum up, the features that we decide to use for the final classifier are the best fitting period $P_{\rm best}$, the de-reddened colour index $g - r$, the amplitudes $A_{k, j}$ for $j = \{1, 2, 3\}$ and the relative phases $\phi_{k, 21}$ and $\phi_{k, 31}$ in the $g$ and $r$ bands, and $\delta \chi^2_{\rm F, C}$, summarized in Table~\ref{tab:features_rfc2}. 
With these features and the label from the SOS \gaia DR2 RRL catalogue, we have prepared all the ingredients for the final classification of the RRL stars among the dataset of 3 million sources.

% --------------------------------------------------
\subsubsection{Random forest classifier II}
\label{subsubsec:method_rfc2}
% --------------------------------------------------

To carry out the last step of the binary classification task, we again utilize the random forest classifier in \code{Scikit-Learn} \citep{scikit-learn} and describe the detail of the process below. 
First, we partition the dataset of 3 million sources into two subsets, the high-quality set and the low-quality set, due to the incompleteness of the SOS \gaia DR2 RRLs in the low galactic latitude areas and the low \gaia epoch areas.
Based on \textsc{HEALPix} \citep{2005ApJ...622..759G} pixels with $\texttt{nside} = 128$, if a source is at the pixel with the galactic latitude at $|b| > 10^\circ$ and at the pixel with a number of \gaia epochs larger than the global mean of 250, we assign the source to the high-quality set, otherwise, it goes into the low-quality set. 
The reasoning for this partition is to only train models in the following steps using the high-quality set because the incompleteness of the SOS \gaia DR2 RRLs on the \textsc{HEALPix} pixels that do not satisfy the above criteria is expected to cause some miss-labelled samples in the low-quality set.

For the high-quality set of 1,273,760 sources, we randomly shuffle the rows and partition the set into 10 subsets such that we can perform a 10-fold cross-validation prediction by training 10 classifiers. 
That is, we train a random forest classifier using all the sources that are not in the $k^{\rm th}$ subset as the training data to predict the probability of each source in the $k^{\rm th}$ subset for $k = \{1, 2, ... , 10\}$ to be a RRL. 
For the low-quality set of 1,767,917 sources, we use the entire high-quality set as the training data to train a random forest classifier and predict the probability of each source in the low-quality set to be a RRL. 
For each random forest classifier, the classifier parameters are the same as the ones used for the random forest classifier I in Section~\ref{subsubsec:method_rfc1}. 
In the end, we concatenate both sets back together to a single set and thus have the predicted probability for each of the 3 million sources being a RRL from the result of the final random forest classification.

% --------------------------------------------------
\subsubsection{Determination of the probability threshold}
\label{subsubsec:result_determine_prob_thre}
% --------------------------------------------------

% Fig ----------------------------------------------
\begin{figure}
    \includegraphics[width=\columnwidth]{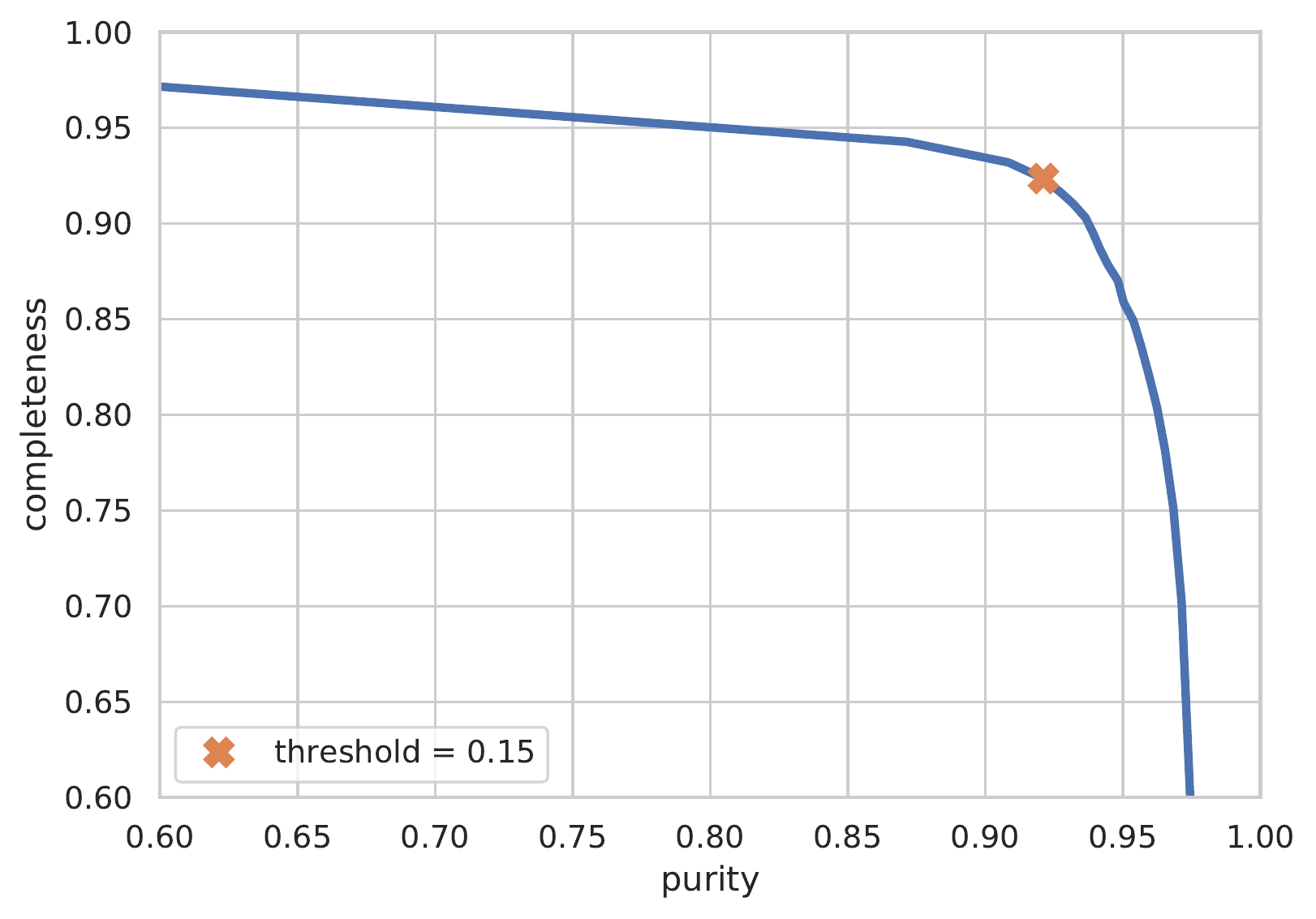}
    \caption{Completeness versus purity of the predicted RRL catalogue with probability thresholds between 0 and 1. The orange mark shows the probability threshold of 0.15.} 
    \label{fig:purity_completeness}
\end{figure}
% Fig ----------------------------------------------

Using the predicted probability for each source being a RRL in the dataset of 3 million sources from Section~\ref{subsubsec:method_rfc2}, we investigate the completeness and purity for different probability thresholds to determine the threshold for our RRL catalogue. 
Given a probability threshold, to compute the completeness and purity of the predicted RRLs, we compare our predicted RRLs to the SOS \gaia DR2 RRL samples in the high galactic latitude areas with $|b| > 10^\circ$ and in the high \gaia epoch areas with the number of \gaia epochs $> 250$ as the high-quality set explained in Section~\ref{subsec:method_final_classification}. 
The reason for applying these two conditions to the calculation of completeness and purity is that the SOS \gaia DR2 RRL samples in these areas are supposed to be more complete compared to the other areas.  
We show the completeness and purity of the predicted RRLs for different probability thresholds in Figure~\ref{fig:purity_completeness}, choosing the probability threshold of 0.15 which maximizes the $F_1$ score\footnote{We note that completeness and purity are the synonyms of recall and precision respectively.} defined as
\begin{equation}
    F_1 = 2 \cdot \frac{{\rm completeness} \cdot {\rm purity}}{{\rm completeness} + {\rm purity}}
\end{equation}
as the orange cross mark shows.  
The probability threshold of 0.15 results in a RRL catalogue of 71,755 predicted RRLs with 0.92 purity and 0.92 completeness, which contains 39,502 out of the original labels of 48,365 SOS \gaia DR2 RRLs.

\section{The RRL catalogue}
\label{sec:result}

% --------------------------------------------------
\subsection{Overview of the catalogue}
\label{subsec:result_overview_catalogue}
% --------------------------------------------------

% Tab ----------------------------------------------
\begin{table}
\caption{The description of our catalogue of 71,755 RRLs. Note that $k = g, r, i$ band in ZTF in the description.}
\label{tab:rrl_cat}
\begin{tabular}{ll}
column                               & description                                                                         \\
\hline
\texttt{objid}                       & ZTF DR3 \texttt{objid}                                                              \\
\texttt{source\_id}                  & \gaia EDR3 \texttt{source\_id}                                                      \\
\texttt{ra}                          & right ascension [deg]                                                               \\
\texttt{dec}                         & declination [deg]                                                                   \\
\texttt{prob\_rrl}                   & predicted probability for being a RRL                                               \\
\texttt{best\_period}                & best fitting period [day]                                                           \\
\texttt{amp\_1\_$k$}                 & $A_{k, 1}$, first Fourier amplitudes [mag]                                          \\
\texttt{amp\_2\_$k$}                 & $A_{k, 2}$, second Fourier amplitudes [mag]                                         \\
\texttt{amp\_3\_$k$}                 & $A_{k, 3}$, third Fourier amplitudes [mag]                                          \\
\texttt{phi\_1\_$k$}                 & $A_{k, 1}$, first Fourier phases [rad]                                              \\
\texttt{phi\_2\_$k$}                 & $A_{k, 1}$, second Fourier phases [rad]                                             \\
\texttt{phi\_3\_$k$}                 & $A_{k, 1}$, third Fourier phases [rad]                                              \\
\texttt{mean\_$k$}                   & $A_{k, 0}$, mean $k$-band magnitude [mag]                                           \\
\texttt{ngooddet\_$k$}               & number of ZTF epochs         \\
\texttt{phot\_g\_mean\_mag}          & \gaia EDR3 mean G magnitude [mag]                                                    \\
\texttt{ebv}                         & $E(B-V)$ [mag]                                                                      \\
\texttt{distance}                    & heliocentric distance [pc]                                                          \\
\hline
\end{tabular}
\end{table}
% Tab ----------------------------------------------

% Tab ----------------------------------------------
\begin{table*}
\caption{A snippet of the machine-readable table for the RRL catalogue (split into three parts below due to space limitation). The detailed description of the columns is in Table~\ref{tab:rrl_cat}.}
\label{tab:rrl_cat_snippet}
    % table 1
    \begin{subtable}[t]{2\columnwidth}
    \centering
    \begin{tabular}[t]{rrrrrrrr}
        \texttt{objid}                      &            
        \texttt{source\_id}                 &       
        \texttt{ra}                         &        
        \texttt{dec}                        &  
        \texttt{prob\_rrl}                  &  
        \texttt{best\_period}               &       
        \texttt{ebv}                        &      
        \texttt{distance}                   \\
        
                                            &
                                            &
        [deg]                               &
        [deg]                               &
                                            &
        [day]                               &
        [mag]                               &
        [pc]                                \\
        \hline
        245101100001850 &  2323207596351730304 &  4.34881 & -26.732536 &      0.95 &     0.621282 &  0.017800 &  35202.500000 \\
        245101200001823 &  2323151181956812672 &  3.44762 & -26.736970 &      0.89 &     0.363568 &  0.022338 &  20908.599609 \\
    \end{tabular}
    \end{subtable}

\bigskip 

    % table 2
    \begin{subtable}[t]{2\columnwidth}
    \centering
    \begin{tabular}[t]{rrrrrrr}
        \texttt{phot\_g\_mean\_mag}     &  
        \texttt{ngooddet\_r}            &  
        \texttt{ngooddet\_g}            &  
        \texttt{ngooddet\_i}            &     
        \texttt{mean\_r}                &     
        \texttt{mean\_g}                &  
        \texttt{mean\_i}                \\

        [mag]                                &
                                             &
                                             &
                                             &
        [mag]                                &
        [mag]                                &
        [mag]                                \\
        
        \hline
        18.278099 &          76 &          74 &           0 &  18.263773 &  18.448895 &         \\
        17.573000 &          81 &          80 &           0 &  17.580330 &  17.662470 &         \\
    \end{tabular}
    \end{subtable}

\bigskip 

    % table 3
    \begin{subtable}[t]{2\columnwidth}
    \centering
    \begin{tabular}[t]{rrrrrrrrr}
        \texttt{amp\_1\_r} &   
        \texttt{amp\_1\_g} &  
        \texttt{amp\_1\_i} &   
        \texttt{amp\_2\_r} &   
        \texttt{amp\_2\_g} &  
        \texttt{amp\_2\_i} &   
        \texttt{amp\_3\_r} &   
        \texttt{amp\_3\_g} &  
        \texttt{amp\_3\_i} \\
        
        [mag]                                &
        [mag]                                &
        [mag]                                &
        [mag]                                &
        [mag]                                &
        [mag]                                &
        [mag]                                &
        [mag]                                &
        [mag]                                \\
        
        \hline
        0.310712 &  0.429619 &          &  0.136991 &  0.186742 &          &  0.077740 &  0.111422 &          \\
        0.214430 &  0.310684 &          &  0.025369 &  0.049442 &          &  0.029519 &  0.016883 &          \\
    \end{tabular}
    \end{subtable}

\bigskip 

    % table 4
    \begin{subtable}[t]{2\columnwidth}
    \centering
    \begin{tabular}[t]{rrrrrrrrr}
        \texttt{phi\_1\_r} &   
        \texttt{phi\_1\_g} &  
        \texttt{phi\_1\_i} &   
        \texttt{phi\_2\_r} &   
        \texttt{phi\_2\_g} &  
        \texttt{phi\_2\_i} &   
        \texttt{phi\_3\_r} &   
        \texttt{phi\_3\_g} &  
        \texttt{phi\_3\_i} \\
        
        [rad]                                &
        [rad]                                &
        [rad]                                &
        [rad]                                &
        [rad]                                &
        [rad]                                &
        [rad]                                &
        [rad]                                &
        [rad]                                \\
                                             
        \hline
        -0.420907 & -0.554002 &          &  1.119307 &  1.172803 &          &  2.935552 &  2.603256 &          \\
         2.610607 &  2.676576 &          &  0.784964 &  0.691132 &          & -1.650670 & -1.853666 &          \\
    \end{tabular}
    \end{subtable}
    
\end{table*}
% Tab ----------------------------------------------

% Fig ----------------------------------------------
\begin{figure*}
    \includegraphics[width=2\columnwidth]{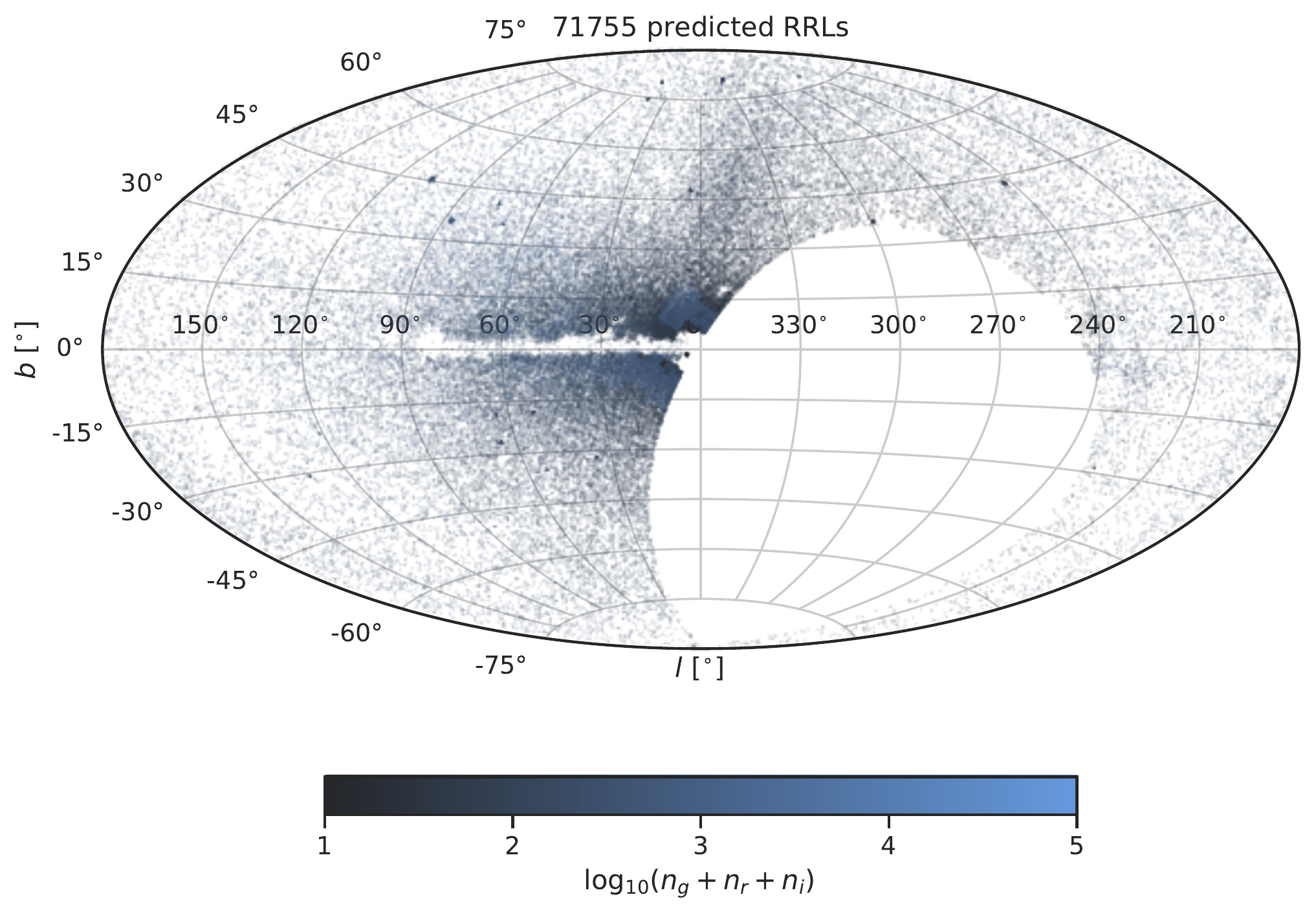}
    \caption{The distribution of our 71,755 RRLs in the Galactic coordinates, color-coded by the total number of ZTf observation epochs in the $gri$ bands. There are some visible stripes associated with the ZTF fields along declination. 
    } 
    \label{fig:cat_rrl_galactic_distribution}
\end{figure*}
% Fig ----------------------------------------------

% Fig ----------------------------------------------
\begin{figure}
    \includegraphics[width=\columnwidth]{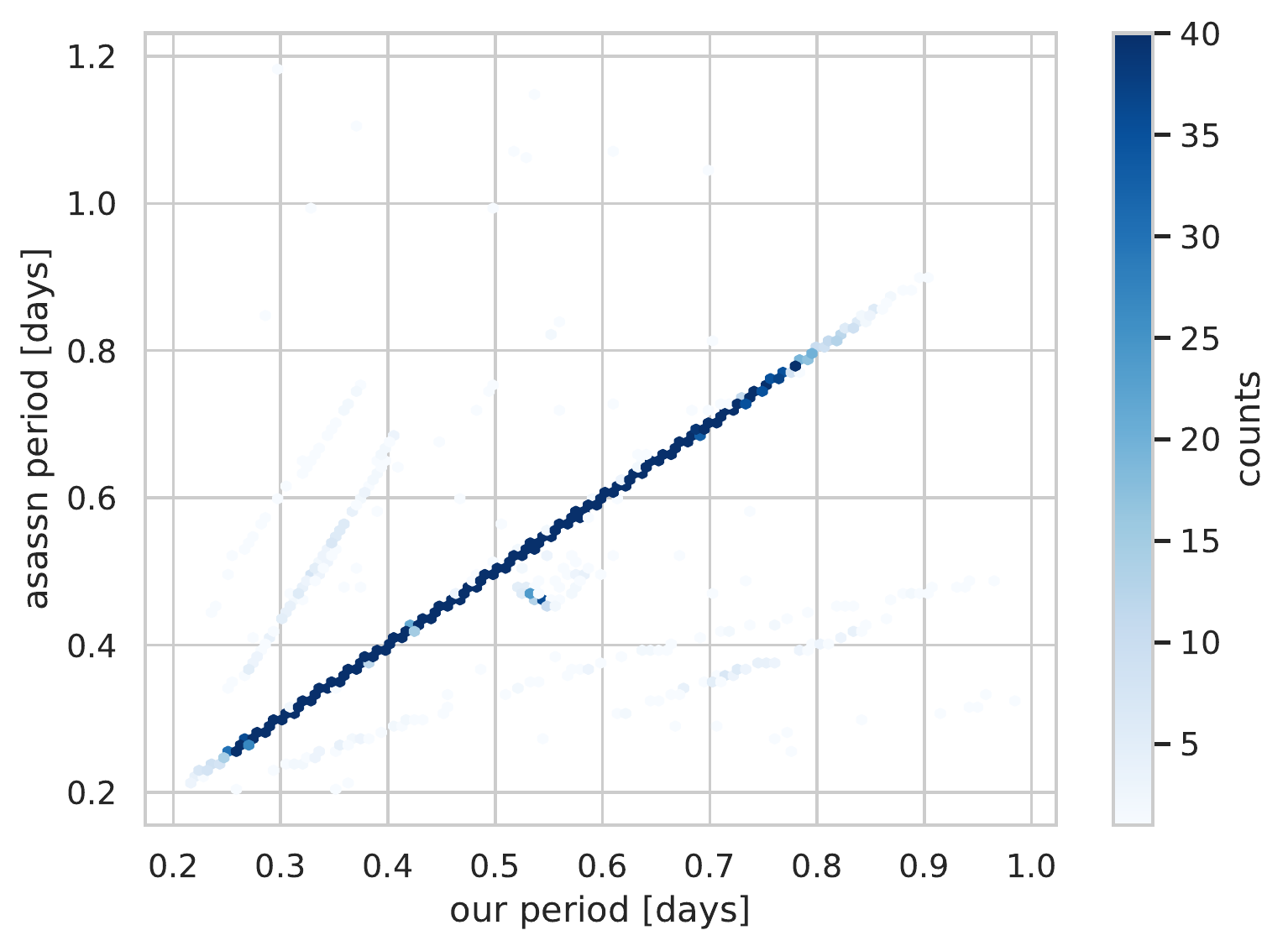}
    \caption{Our best fitting period $P_{\rm best}$ versus the period provided by the ASAS-SN catalogue \citep{2020yCat.2366....0J} for the common 18,854 RRLs in both datasets.} 
    \label{fig:rrl_period_comparison}
\end{figure}
% Fig ----------------------------------------------

% Fig ----------------------------------------------
\begin{figure}
    \includegraphics[width=\columnwidth]{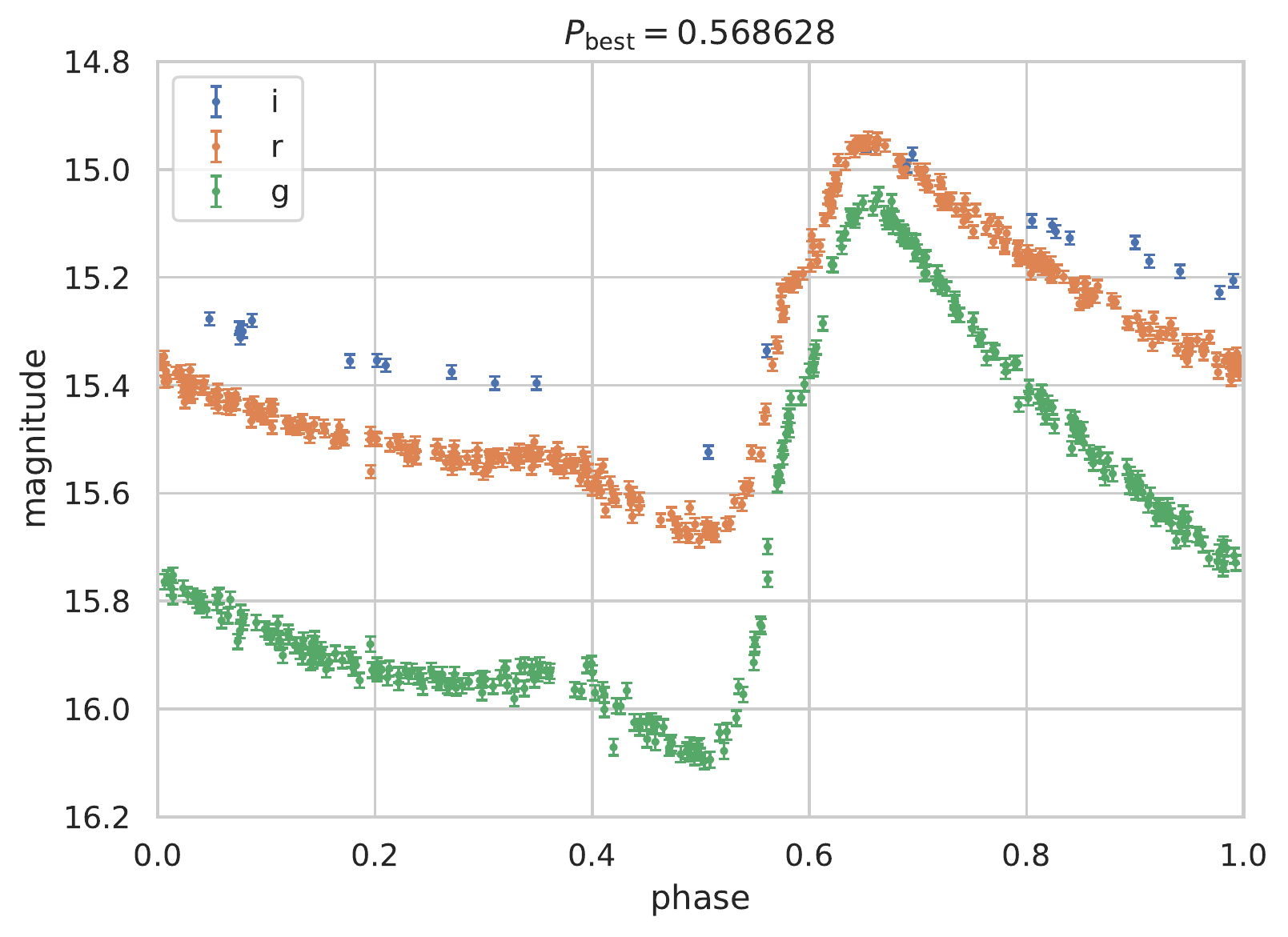}
    \caption{An example of RRL ZTF light curves folded by its best period $P_{\rm best}$, whose \gaia EDR3 $\texttt{source\_id} = 2294134898301488640$.} 
    \label{fig:example_light_curve}
\end{figure}
% Fig ----------------------------------------------

In this section, we give an overview of the RRL catalogue produced by the pipeline described in Section~\ref{sec:method_pipeline}.
Covering the Northern sky, this catalogue containing 71,755 RRLs in the joint set of \gaia EDR3 and ZTF DR3 will be the main RRL catalogue of the paper. 
A detailed description of the catalogue contents is provided in Table~\ref{tab:rrl_cat}. The catalogue is released in electronic form with the paper at \href{https://doi.org/10.5281/zenodo.5774018}{DOI 10.5281/zenodo.5774017} \citep{huang_kuan_wei_2021_5774018} with a short snippet of the table provided in Table~\ref{tab:rrl_cat_snippet}.

To evaluate the heliocentric distances in the catalogue, we first derive the absolute magnitudes of the RRLs according to the PS1 period-luminosity relations in \citet{2017AJ....153..204S} assuming a halo metallicity of $\left[ {\rm Fe} / {\rm H} \right] = -1.5$ \citep{2008ApJ...684..287I}
\begin{equation}
\label{eq:ps1_PL_relation}
\begin{split}
    M_g &= -1.7  \log_{10} \left( \frac{P_{\rm best}}{0.6} \right) + 0.69 \\
    M_r &= -1.6  \log_{10} \left( \frac{P_{\rm best}}{0.6} \right) + 0.51 \\
    M_i &= -1.77 \log_{10} \left( \frac{P_{\rm best}}{0.6} \right) + 0.46.
\end{split}
\end{equation}
Together with the mean ZTF magnitudes as the zeroth-order fitted Fourier amplitude $A_{k, 0}$ for $k=g, r, i$ corrected by the extinction in \citet{2011ApJ...737..103S}, we evaluate the distance moduli $\mu_k$ as
\begin{equation}
\label{eq:apparent_mag_ext_corrected}
\begin{split}
    \mu_g &= A_{g, 0} - 3.17 E(B-V) - M_g  \\ 
    \mu_r &= A_{r, 0} - 2.27 E(B-V) - M_r  \\ 
    \mu_i &= A_{i, 0} - 1.68 E(B-V) - M_i  
\end{split}
\end{equation}
and then derive the heliocentric distance by averaging the distance moduli.

As a first look at the catalogue, we show the sky distribution of the 71,755 predicted RRLs in the Galactic coordinates in Figure~\ref{fig:cat_rrl_galactic_distribution}, observing the Galactic halo and the Sagittarius Stream despite the lack of coverage of the Southern sky. 
Compared to the SOS \gaia DR2 RRLs which serves as the label in our classification pipeline, there are several facts about our RRL catalogue which are worth noting. 
Our RRL catalogue contains more sources than the 48,365 SOS \gaia RRL samples in the Northern sky coverage with the completeness of 0.92 and purity of 0.92 globally. 
Colour-coded by the total ZTF observation epochs, Figure~\ref{fig:cat_rrl_galactic_distribution} shows that our RRL catalogue is more complete in the areas where \gaia suffers incompleteness due to its scanning trajectory as the patches with fewer RRLs shown in Figure~\ref{fig:gaia_rrls_n_epochs_pos}, and in the low galactic latitude areas, e.g. $3^\circ < |b| < 10^\circ$.

To show the robustness of our fitting period, we compare our best-fitting periods to the periods provided in the ASAS-SN catalogue \citep{2020yCat.2366....0J}, for the 18,854 RRLs that are in both catalogues by matching the \gaia EDR3 \texttt{source\_id} provided in both catalogues. 
The reason to choose the ASAS-SN catalogue to compare with is due to its high number of epochs \citep[each ASAS-SN field in the V-band has roughly 100 -- 600 epochs][]{2018RNAAS...2...18J} and thus its reliable period determination.
Figure~\ref{fig:rrl_period_comparison} shows the alignment on the one-to-one line on the period plane and indicates the goodness of our period fitting result. 
We note that 97\% of the 18,854 RRLs have a period percentage difference smaller than 0.1\%, though several objects suffer the aliasing period issue during the Fourier fitting process \citep{1976Ap&SS..39..447L, 1982ApJ...263..835S, 2018ApJS..236...16V}.
Moreover in Figure~\ref{fig:example_light_curve}, we display an example of ZTF light curves from our RRL catalogue in the $gri$ bands, folded by its best-fitting period $P_{\rm best}$. 
Demonstrating a typical shape of a folded RRL light curve, this furthermore shows the robustness of our Fourier Series fitting described in Section~\ref{subsubsec:method_feature_classification} and the resulting period in the catalogue.

To further investigate the RRL catalogue, we will look into the completeness of the catalogue in Section~\ref{subsec:result_completeness}, compare the catalogue with other existing catalogues in Section~\ref{subsec:result_comparison_other_cats}, and study the Galactic halo profile in Section~\ref{subsec:result_Galactic_halo_profile}

% --------------------------------------------------
\subsection{Completeness of the catalogue}
\label{subsec:result_completeness}
% --------------------------------------------------

% Fig ----------------------------------------------
\begin{figure}
    \includegraphics[width=\columnwidth]{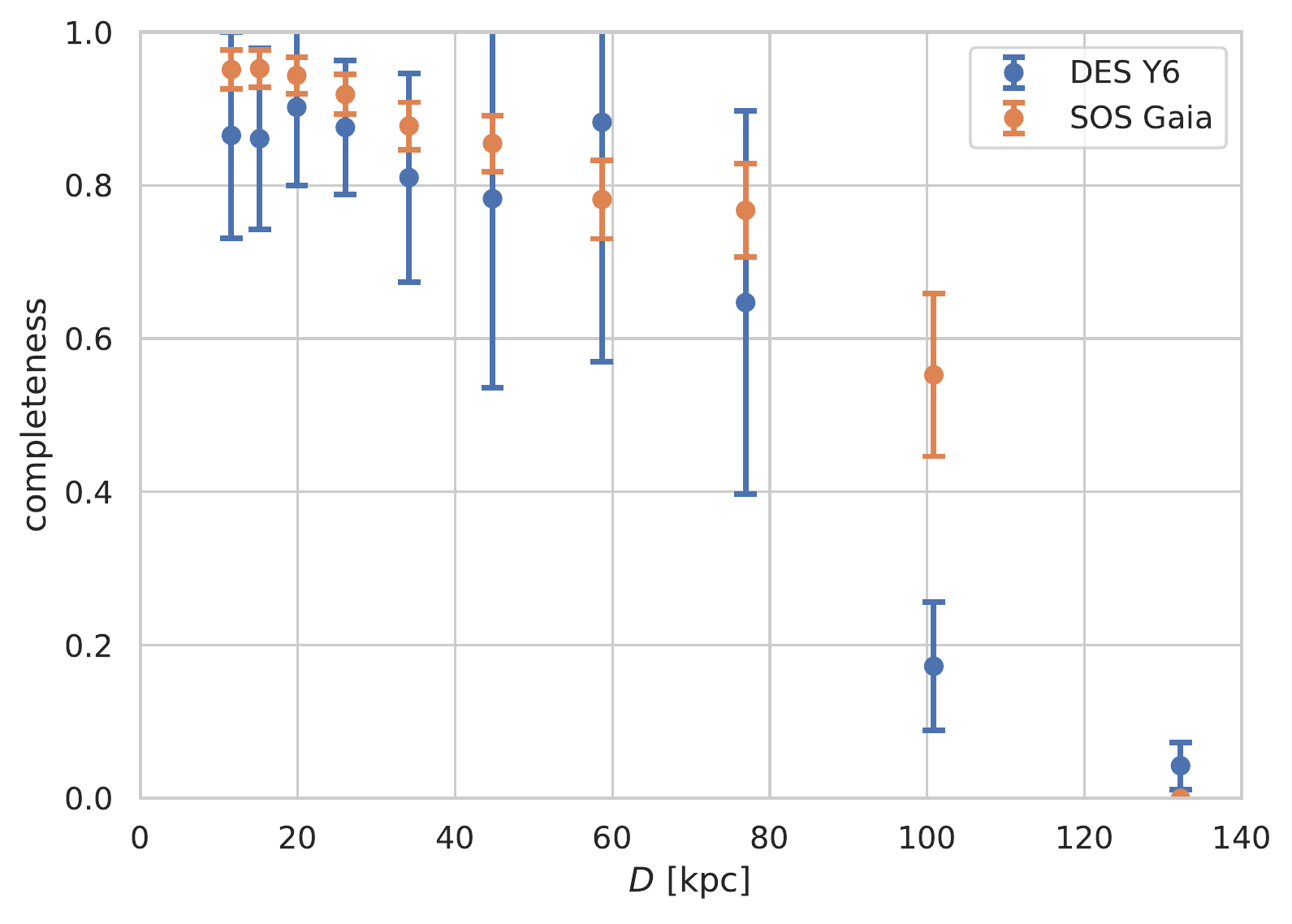}
    \caption{The completeness of our RRL catalogue as a function of heliocentric distance compared to the SOS \gaia DR2 RRL catalogue and the DES Y6 RRL catalogue.} 
    \label{fig:cat_rrl_completeness_distance}
\end{figure}
% Fig ----------------------------------------------

% Fig ----------------------------------------------
\begin{figure*}
    \includegraphics[width=2\columnwidth]{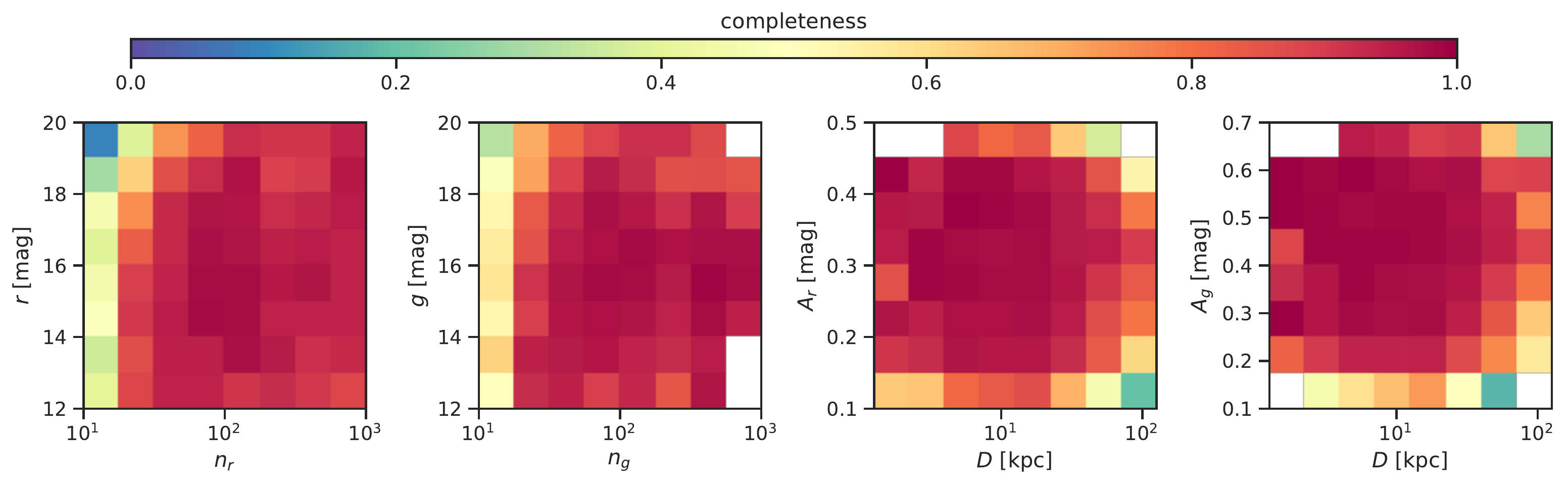}
    \caption{\textbf{Left} and \textbf{middle-left}: the completeness as functions of the mean magnitudes $r$ and $g$ and the ZTF numbers of epochs in $r$ and $g$ bands $n_r$ and $n_g$. \textbf{Middle-right} and \textbf{right}: the completeness as functions of the amplitudes $A_r$ and $A_g$ and the heliocentric distance $D$. 
    } 
    \label{fig:cat_rrl_completeness_different_quans}
\end{figure*}
% Fig ----------------------------------------------

As mentioned in Section~\ref{subsubsec:result_determine_prob_thre}, our RRL catalogue has overall completeness of 0.92 compared to the SOS \gaia DR2 RRLs grouped by the \textsc{HEALPix} pixels with $\texttt{nside} = 128$ with the number of \gaia epochs $> 250$ globally. 
In this section, we will look into the completeness in more detail and we begin by investigating the completeness as a function of heliocentric distance in Figure~\ref{fig:cat_rrl_completeness_distance}. 
Using the SOS \gaia DR2 RRLs grouped by the same \textsc{HEALPix} pixels to compute the completeness in heliocentric distance bins, we find that the completeness is higher than $\sim 0.8$ at the regions with distance smaller than 80 kpc, is roughly 0.5 at 100 kpc, and drops drastically to 0 at 130 kpc. 
We note that the most distant RRL in our catalogue is at a distance of 132~kpc. 
Thanks to the deeper RRL catalogue from DES Y6 with the most distant RRL at $\sim 300$ kpc \citep{2021ApJ...911..109S}, we cross-match the closest RRL within one arcsec at the areas above $-20^\circ$ declination and evaluate the completeness, finding that the completeness is consistent with the one compared to the SOS \gaia RRLs for distance smaller than 80 kpc.  
However, at distance larger than 100 kpc, the completeness drastically drops to 0.2 and then 0.

We move on to investigate the influence of several quantities on the completeness of our RRL catalogue, including the distance, the amplitudes, the magnitudes, and the numbers of epochs, again utilizing the SOS \gaia DR2 RRLs on the \textsc{HEALPix} pixels with $\texttt{nside} = 128$ with the number of \gaia epochs $> 250$. 
In the left and the middle-left panels of Figure~\ref{fig:cat_rrl_completeness_different_quans}, we show the completeness as a function of $r$ and $n_r$ and that of $g$ and $n_g$ respectively, where $r$ and $g$ are the mean magnitudes corrected by the extinction and $n_r$ and $n_g$ are the numbers of ZTF detection with $\texttt{catflags} < 32768$ in $r$ and $g$ bands. 
We find that the completeness is lower when there is less detection for a source given a magnitude and when the luminosity is fainter given a number of detection. 
The middle-right and the right panels show the completeness as a function of the $r$-band amplitude $A_r$ and the heliocentric distance $D$ and that of $g$-band amplitude $A_g$ and $D$ respectively.
The amplitudes $A_r$ and $A_g$ defined from the combination of multiple Fourier terms
\begin{equation}
\label{eq:amp_r_g}
\begin{split}
    A_r &= \sqrt{ A_{r, 1}^2 + A_{r, 2}^2 + A_{r, 3}^2 } \\
    A_g &= \sqrt{ A_{g, 1}^2 + A_{g, 2}^2 + A_{g, 3}^2 }
\end{split}
\end{equation}
are the best fitting amplitudes from the third-order Fourier Series in the $g$ and $r$ bands.
We find that the completeness gradually decreases as the distance increases given an amplitude, meaning that our catalogue is less complete at more distant regions, which is consistent with Figure~\ref{fig:cat_rrl_completeness_distance}. 
When given a distance, the completeness drops faster at the small-amplitude ends than at the large-amplitude end.

% --------------------------------------------------
\subsection{Comparison with other catalogues}
\label{subsec:result_comparison_other_cats}
% --------------------------------------------------

% Fig ----------------------------------------------
\begin{figure*}
    \includegraphics[width=2\columnwidth]{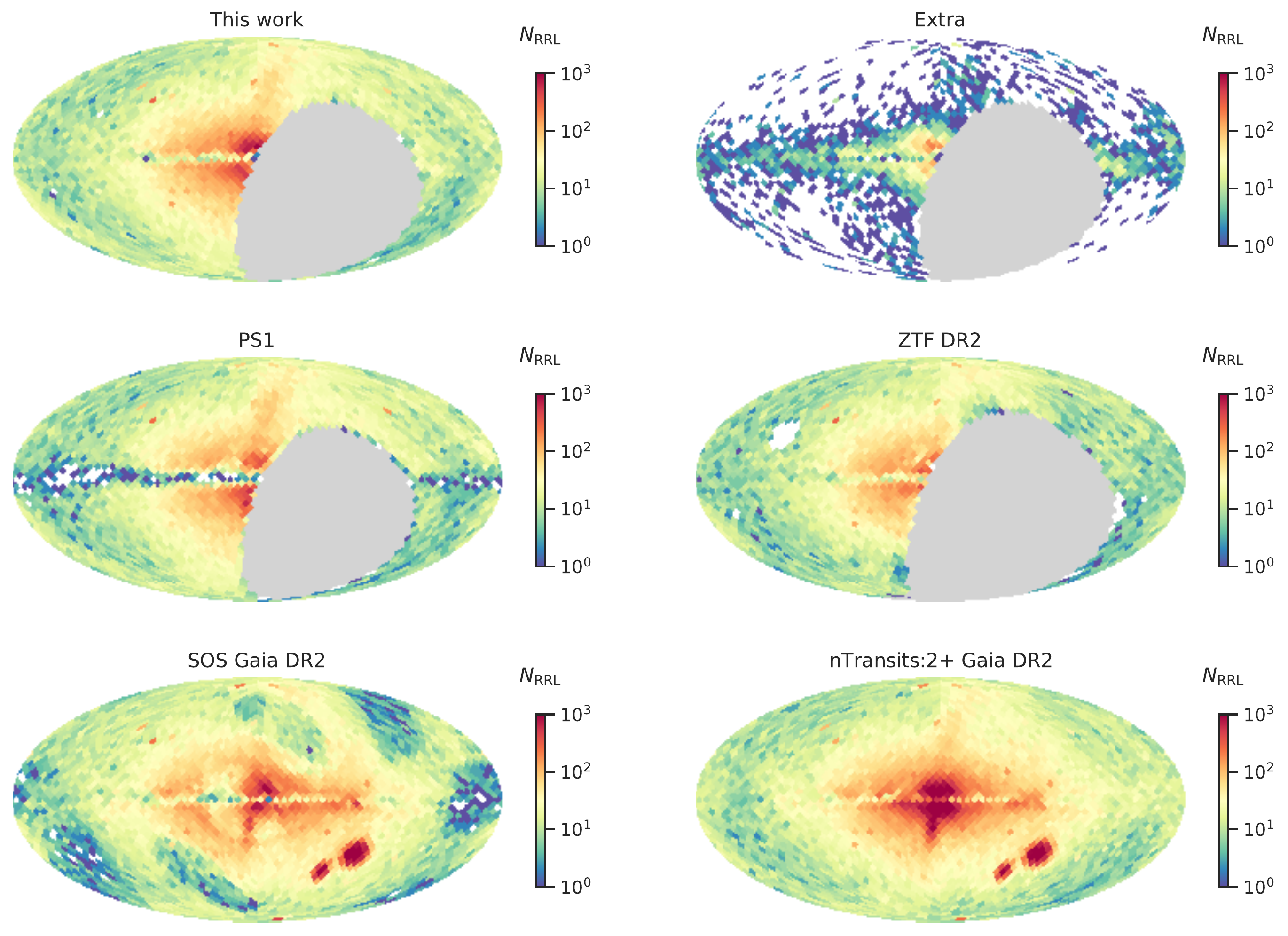}
    \caption{The RRL distributions in the Galactic coordinate of the catalogue from this work, the PS1 catalogue \citep{2017AJ....153..204S}, the ZTF DR2 catalogue \citep{2020ApJS..249...18C}, the SOS \gaia DR2 catalogue \citep{2019A&A...622A..60C}, and the nTransits:2+ \gaia DR2 catalogue \citep{2018A&A...618A..30H}, colour-coded by the number of RRLs on each grid $N_{\rm RRL}$. The top-right panel illustrates the extra RRLs from this work that are not in any external catalogues mentioned in Section~\ref{subsec:result_comparison_other_cats}.} 
    \label{fig:cats_rrl_comparison_galactic_distribution}
\end{figure*}
% Fig ----------------------------------------------

% Tab ----------------------------------------------
\begin{table}
\caption{The completeness of our catalogue compared to some external RRL catalogues. For each catalogue, we apply the selections of declination $> -20^\circ$, $|b| > 10^\circ$, and magnitude between 15 and 20. 
After the selections, $N$ is the number of RRLs in the external catalogues, and $N_{\rm x}$ is the number of RRLs from each external catalogue that have a match in our catalogue.}
\label{tab:rrl_external_cats_comparison}
\begin{tabular}{lllll}
    catalogue      & $N$ & $N_{\rm x}$ & $N_{\rm x} / N$      & reference(s) \\
    
    \hline
    ZTF DR2     & 28883   & 27993    & 0.96        & \citet{2020ApJS..249...18C}    \\
    DES Y6      & 769     & 665      & 0.86        & \citet{2021ApJ...911..109S}    \\
    ASAS-SN     & 12765   & 10704    & 0.83        & \citet{2018RNAAS...2...18J}    \\
    PS1         & 32045   & 25862    & 0.80        & \citet{2017AJ....153..204S}    \\
    SOS         & 30002   & 24090    & 0.80        & \citet{2019AA...622A..60C}     \\
    OGLE        & 701     & 567      & 0.80        & \citet{2019AcA....69..321S}    \\
    CRTS        & 6917    & 5473     & 0.79        & \citet{2014ApJS..213....9D}    \\
    % NSVS        & 8       & 5        & 0.62        & \citet{2006MNRAS.368.1757W}    \\
    
    \hline
\end{tabular}
\end{table}
% Tab ----------------------------------------------

We start this section by comparing our RRL catalogue to several recent RRL catalogues covering the entire Northern sky. 
Figure~\ref{fig:cats_rrl_comparison_galactic_distribution} shows the RRL distributions of different catalogues in the Galactic coordinate, colour-coded by the number of RRLs $N_{\rm RRL}$ on each \textsc{HEALPix} pixel of $\texttt{nside} = 16$. 
The catalogues plotted are the RRL catalogue from this work, the PS1 catalogue \citep{2017AJ....153..204S}, the ZTF DR2 catalogue \citep{2020ApJS..249...18C}, the SOS \gaia DR2 catalogue \citep{2019A&A...622A..60C}, and the nTransits:2+ \gaia DR2 catalogue \citep{2018A&A...618A..30H}. 
In particular, we apply the score thresholds of 0.8 and 0.55 for types ab and c RRLs according to \citet{2017AJ....153..204S} when utilizing the PS1 catalogue.

Overall, our catalogue, the PS1 catalogue, and the nTransits:2+ \gaia DR2 catalogue illustrate the Galactic halo and the Sagittarius Stream better than the ZTF DR2 catalogue and the SOS \gaia DR2 catalogue do.  
Even though the nTransits:2+ \gaia DR2 catalogue covers the whole sky, it is generally more contaminated than the SOS RRL catalogue \citep[see][for more detail]{2018A&A...618A..30H, 2019A&A...622A..60C} and it does not provide periods and light curve fits for the RRL samples. 
Compared to the SOS \gaia DR2 catalogue which serves as our training label, our RRL catalogue outperforms at the incomplete areas caused by the \gaia scanning trajectory and has more RRLs in the Northern sky coverage. 
Compared to the PS1 catalogue of 61,144 RRLs, our catalogue has more RRL samples, especially around the Galactic halo, and covers the low galactic latitude areas better. 
Compared to the ZTF DR2 catalogue of 46,358 RRLs, our catalogue has more RRLs globally, especially near the Galactic halo and the Sagittarius Stream, and tends to have more numbers of observed epochs due to the usage of ZTF DR3.

Besides the above five catalogues covering the entire Northern sky, we also compare our catalogue to other existing RRL catalogues, including the DES Y6 catalogue \citep{2021ApJ...911..109S}, the CRTS catalogue \citep{2014ApJS..213....9D}, the ASAS-SN catalogue \citep{2018RNAAS...2...18J}, the OGLE catalogue \citep{2019AcA....69..321S}, and the NSVS catalogue \citep{2006MNRAS.368.1757W}. 
For the comparison, we apply three selections on every catalogue, the selection of declination $> -20^\circ$ due to the sky coverage of the ZTF survey, the selection of $|b| > 10^\circ$ to exclude the region near the Galactic disc, and the selection of magnitude between 15 and 20 based on our RRL magnitude distribution because the depth of the catalogues varies. 
After the selections, we count the number of RRLs in each catalogue $N$, amongst them we count the number of RRLs from each catalogue that have a match in our table as $N_{\rm x}$, and from that we calculate the overall completeness of our catalogue as  $N_{\rm x} / N$.
The cross-matching is done by selecting the closest objects based on the angular separation within 1 arcsec for most of the catalogues, except for the CRTS and ASAS-SN catalogues. 
When cross-matching the CRTS catalogue to our catalogue, we use the angular separation of 2.5 arcsec as it is the pixel size for CRTS \citep{2009ApJ...696..870D}. 
For the ASAS-SN catalogue, it has already provided the \gaia EDR3 \texttt{source\_id}, which our catalogue also provides, so we directly utilize the \texttt{source\_id} to cross-match the two catalogues.

The results of the comparison are summarized in Table~\ref{tab:rrl_external_cats_comparison}. 
We note that there are only 8 stars left in the NSVS catalogue after the selections, so we exclude NSVS from the table. 
Our catalogue achieves high completeness of 96\% compared to the ZTF DR2 catalogue, which is expected to be the highest as these two catalogues are based on the same survey but different data releases. 
For all the other catalogues, DES Y6, ASAS-SN, PS1, \gaia SOS, OGLE, and CRTS, our catalogue has the completeness $\gtrsim 80\%$. 
We note that our catalogue is possibly less complete for distant RRLs, for small-amplitude RRLs, for type c RRLs, or for the RRLs located on the field boundary regions.

We end the section by identifying the extra RRLs from our catalogue when cross-matched with all the external RRL catalogues mentioned in the section. 
In total, we have 6547 extra RRLs, and we visualize them in the top-right panel in Figure~\ref{fig:cats_rrl_comparison_galactic_distribution}. 
This panel indicates the extra RRLs in our catalogue concentrate around the Galactic halo and near the Galactic disk. 
When making this panel, we mask out 844 RRLs with the period within $0.5 \pm 0.01$ days because they are most likely contaminated objects due to the aliasing period issue.

% --------------------------------------------------
\subsection{The Galactic halo profile}
\label{subsec:result_Galactic_halo_profile}
% --------------------------------------------------

% Fig ----------------------------------------------
\begin{figure}
    \includegraphics[width=\columnwidth]{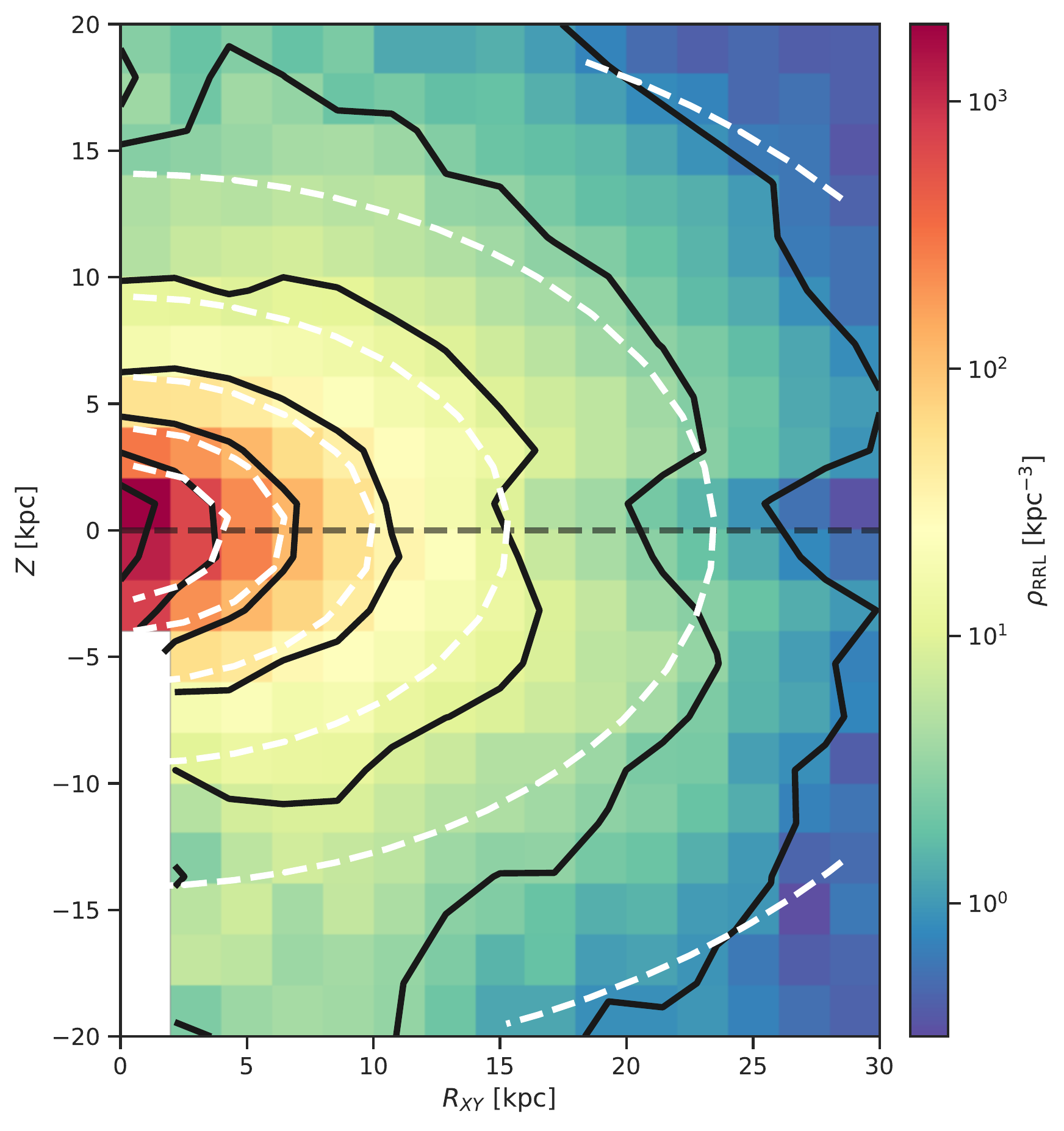}
    \caption{The 2D histogram of the RRL distribution in the cylindrical galactocentric coordinates ($R_{XY}-Z$)
     colour-coded by the RRL number density $\rho_{\rm RRL}$ on each grid. The black curves are the contours of $\rho_{\rm RRL} = 10^{0}, 10^{0.5}, 10^{1}, 10^{1.5}, 10^{2}, 10^{2.5}, 10^{3}$ ${\rm kpc}^{-3}$. The white elliptical contours are the single power law density profile with $q = 0.6$ and power of -2.7 from \citet{2018MNRAS.474.2142I}.} 
    \label{fig:cat_rrl_density_rxy_z}
\end{figure}
% Fig ----------------------------------------------

% Fig ----------------------------------------------
\begin{figure*}
    \includegraphics[width=2\columnwidth]{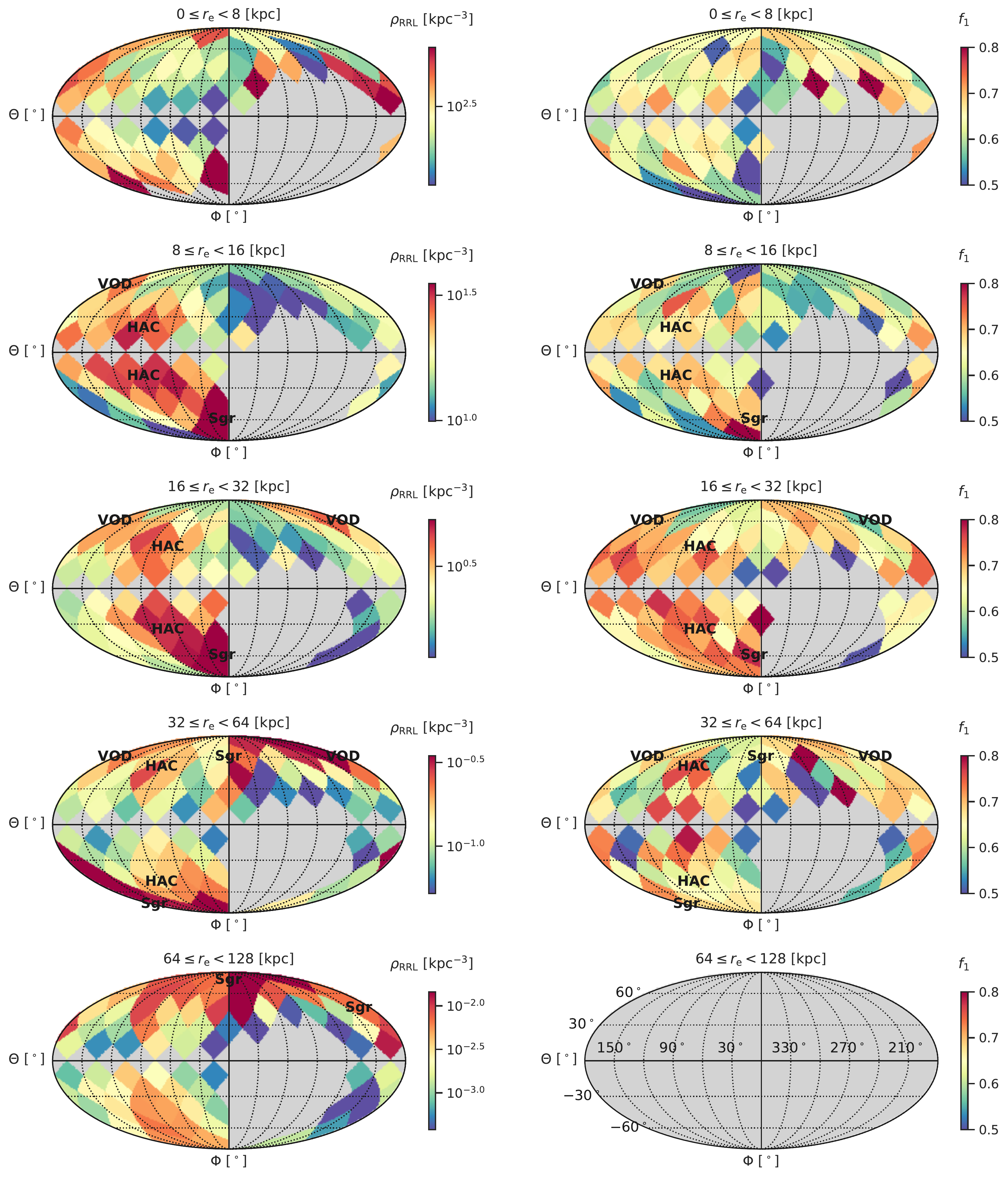}
    \caption{\textbf{Left column}: The RRL number density $\rho_{\rm RRL}$ on spheroidal shells of different elliptical radii $r_{\rm e}$ in the coordinate of the Galactocentric longitude $\Phi$ and latitude $\Theta$. \textbf{Right column}: The Oosterhoff type I fraction $f_1$ on each spheroidal shell in $\Phi$ and $\Theta$. For the grids on each panel, the edges from left to right are $\Phi = 180^\circ, 150^\circ, 120^\circ, 90^\circ, 60^\circ, 30^\circ, 0^\circ, 330^\circ, 300^\circ, 270^\circ, 240^\circ, 210^\circ, 180^\circ$ and from top to bottom are $\Theta = 90^\circ, 60^\circ, 30^\circ, 0^\circ, -30^\circ, -60^\circ, -90^\circ$. The annotations HAC, VOD, and Sgr are the Hercules-Aquila Cloud, the Virgo over-density, and the Sagittarius Stream respectively.} 
    \label{fig:cat_rrl_density_f1_re_slices}
\end{figure*}
% Fig ----------------------------------------------

% Fig ----------------------------------------------
\begin{figure}
    \includegraphics[width=\columnwidth]{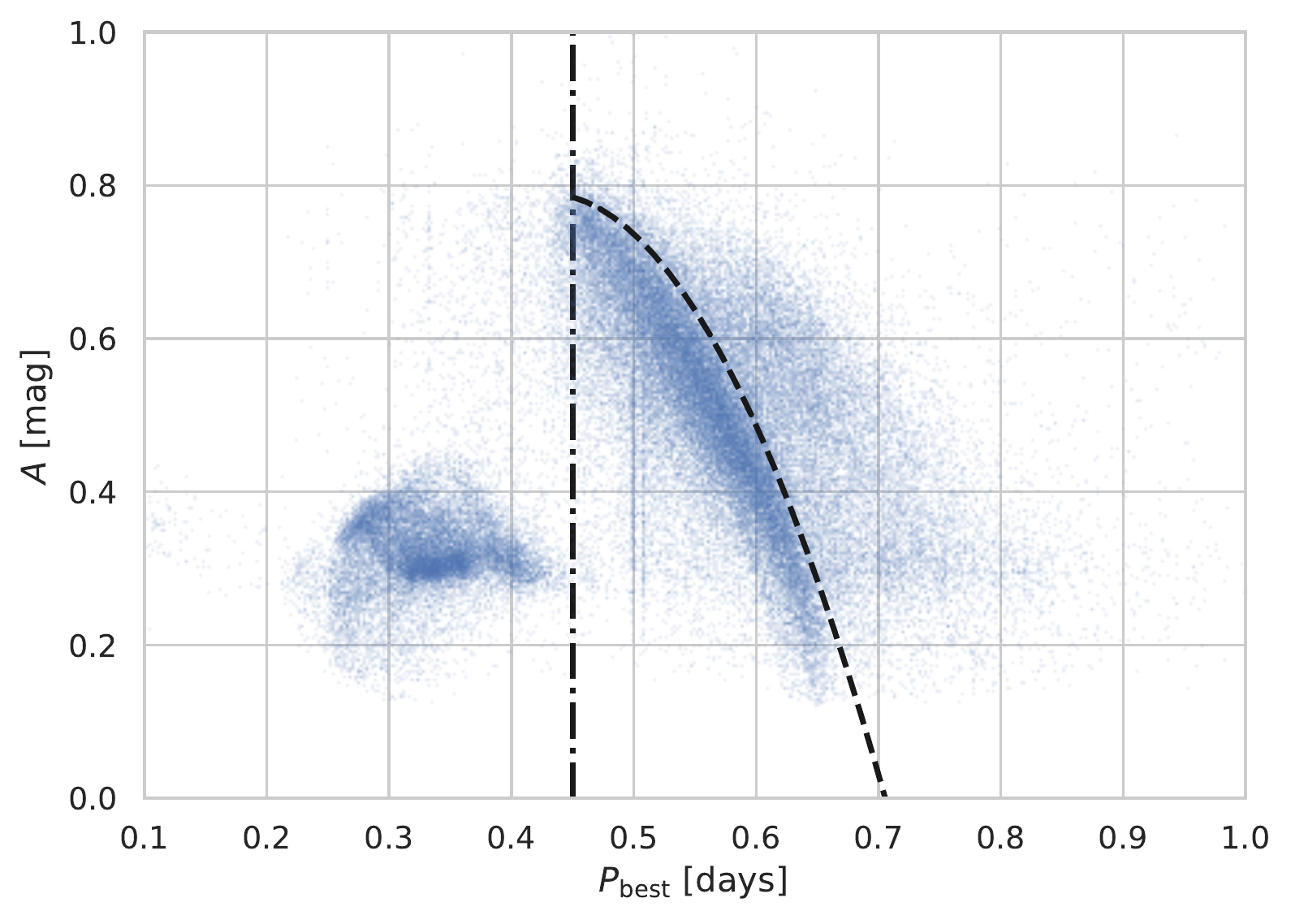}
    \caption{The distribution of detected RRLs on the period-amplitude diagram, where $P_{\rm best}$ is the period and $A = \sqrt{A^2_r + A^2_g}$ is the total amplitude of the best fit in $g$ and $r$ bands. The dash-dotted line of $P_{\rm best} = 0.45$ days is the boundary to roughly separate RRab and RRc stars. The dashed curve is the boundary we adopt to separate Oosterhoff I and II for RRab stars in Equation~\ref{eq:Oo_boundary}.}
    \label{fig:cat_amp_period}
\end{figure}
% Fig ----------------------------------------------

Knowing that our catalogue contains more RRLs around the Galactic halo and near the low Galactic latitude areas compared to the other catalogues in Section~\ref{subsec:result_comparison_other_cats}, we study the Galactic halo profile using our RRL catalogue in the Galactocentric coordinate in this section. 
Focusing on the Galactic halo profile, we mask out the RRLs in the Milky Way dwarf galaxies and globular clusters with declination above $-28^\circ$ due to the coverage of ZTF and with heliocentric distance smaller than 100 kpc due to the completeness of our catalogue. 
This criterion includes 90 globular clusters from \citet{1996AJ....112.1487H, 2010arXiv1012.3224H} and 17 dwarf galaxies of Bootes I and II, Cetus II, Coma Berenices, Draco, Draco II, Sagittarius II, Segue I and II, Sextans I, Triangulum II, Ursa Major I and II, Ursa Minor, and Willman I from \citet{2012AJ....144....4M}, and Bootes III \citep{2018AA...620A.155M} and Virgo I \citep{Homma_2016}.
After the selection, there are 70950 RRLs for the study of the Galactic halo profile in this section.

We briefly lay out the Galactocentric coordinate adopted in this section. 
The right-handed Cartesian coordinate $(X, Y, Z)$ is computed by the Galactic longitude, Galactic latitude, and heliocentric distance $(l, b, D)$ as
\begin{equation} 
\label{eq:cartesian_coordinate}
\begin{split}
    X &= D \cos l \cos b - R_{\sun} \\
    Y &= D \sin l \cos b \\
    Z &= D \sin b 
\end{split}
\end{equation}
where $R_{\sun} = 8$ kpc is the distance between the Galactic Centre and the Sun. 
This coordinate is centred at the Galactic Centre with the Galactic disk on the $(X, Y)$ plane, the $Z$-axis pointing to the north Galactic pole, and the $X$-axis pointing from the Sun at $X = -8$ kpc to the Galactic Centre at $X = 0$ kpc. 
We define the cylindrical radius $R_{XY}$ and the elliptical radius $r_{\rm e}$ as 
\begin{equation} 
\label{eq:Rxy_Rq}
\begin{split}
    &R_{XY} = \sqrt{X^2 + Y^2} \\
    &r_{\rm e} = \sqrt{X^2 + Y^2 + \left( \frac{Z}{q} \right)^2} 
\end{split}
\end{equation}
where the flattening $q \sim 0.6$ for the spheroidal stratification according to literature about the Galactic density profile fitting \citep[e.g.][]{2018MNRAS.474.2142I}.
We further define the Galactocentric longitude $\Phi$ and latitude $\Theta$ as
\begin{equation} 
\label{eq:phi_theta}
\begin{split}
    \Theta &= \arctan \frac{Z}{R_{XY}} \\
    \Phi &= \arctan \frac{Y}{X}.
\end{split}
\end{equation}

To study the density profile of the Galactic halo from different perspectives in the Galactocentric coordinate, we need to evaluate the RRL number density $\rho_{\rm RRL}$ based on our RRL catalogue. 
The calculation of $\rho_{\rm RRL}$ is the number of RRLs per volume, where we take into account two factors when evaluating the volume: the ZTF coverage of declination $> -28^\circ$ and the completeness as a function of ZTF epoch and magnitude. 
Given a grid at $(X, Y, Z)$, we compute its declination and count the grid if it is above $-28^\circ$. 
Based on the position of the grid, we calculate the mean $r$-band epoch utilizing \textsc{HEALPix} with $\texttt{nside} = 8$ for all ZTF sources. 
Besides, knowing the heliocentric distance of the grid and using the $r$-band absolute magnitude of $M_r = 0.54$ mag which maximizes the histogram of $M_r$ of all 71,755 RRLs, we evaluate the $r$-band magnitude of RRLs for the grid.
With the mean $r$-band epoch and the $r$-band magnitude of RRLs for the given grid, we compute the completeness on the grid by interpolating the value from the completeness matrix shown in the left panel of Figure~\ref{fig:cat_rrl_completeness_different_quans}.

To visualize the spheroidal stratification of the Galactic halo density profile and to look for Galactic disk RRLs, we show the RRL number density $\rho_{\rm RRL}$ around the Galactic halo on the $R_{XY}-Z$ plane in Figure~\ref{fig:cat_rrl_density_rxy_z}, assuming the density profile is cylindrically symmetric.
The black contours of $\rho_{\rm RRL} = 10^{0}, 10^{0.5}, 10^{1}, 10^{1.5}, 10^{2}, 10^{2.5}, 10^{3}$ ${\rm kpc}^{-3}$ verify the roughly spheroidal density profile with the flattening $q \sim 0.6$ for $r_{\rm e}$ in Equation~\ref{eq:Rxy_Rq}, as indicated by the white dashed elliptical contours. 
The change of the exponent if modelled by the power-law models, indicated by the distance of any two neighbored contours getting larger as the radius increasing, is consistent with the findings of the single power-law in \citet{2018MNRAS.474.2142I}. 
We note that some recent works have found a break in the radial profile of the halo at the Galactocentric distances of 25-30~kpc \citep[e.g.][]{2018ApJ...855...43M, 2021ApJ...911..109S}.
Despite having more RRLs near the disk compared to other catalogues as discussed in Section~\ref{subsec:result_comparison_other_cats}, our catalogue still lacks some RRLs at the regions near the disk with roughly $|Z| < 2$ kpc, which can be seen at the regions with roughly $|b| < 3^\circ$ in Figure~\ref{fig:cat_rrl_galactic_distribution} as well.

As the Galactic halo stellar density profile is potentially triaxial \citep{2018MNRAS.474.2142I}, to study the substructure in the Galactic halo, we also look at the RRL density distribution in the coordinate of Galactocentric longitude $\Phi$ and latitude $\Theta$ defined in Equation~\ref{eq:phi_theta}. 
The left panels in Figure~\ref{fig:cat_rrl_density_f1_re_slices} illustrate the RRL number density $\rho_{\rm RRL}$ on the spheroidal shells of different elliptical radii $0 < r_{\rm e} < 128$ kpc with the flattening $q = 0.6$, each of which demonstrates the density on the sky view with $\Phi = 180^\circ$ pointing to the Sun and $\Phi$ increasing towards the left in the figure. 
We observe and annotate some known over-densities of the Galactic halo, including the Sagittarius Stream \citep{2017ApJ...850...96H}, the Virgo over-density \citep{2001ApJ...554L..33V, 2002ApJ...569..245N, 2006ApJ...636L..97D, 2008ApJ...673..864J, 2012AJ....143..105B}, and the Hercules-Aquila Cloud \citep{2007ApJ...657L..89B, 2014MNRAS.440..161S, 2018MNRAS.476.3913S}. 
An interesting point from the panels of $16 < r_{\rm e} < 64$ kpc is that the Northern part of the Hercules-Aquila Cloud is very close to the Virgo over-density, where the possible association of the two over-densities has been discussed in recent literature \citep[e.g.][]{2016ApJ...817..135L, 2019MNRAS.482..921S, 2021arXiv210409794B}, as well as the Eridanus–Phoenix over-density which however is not in ZTF coverage. 
It is worth noting that there are over-densities in the Northern and the Southern hemispheres with $\Phi$ roughly from $30^\circ$ to $120^\circ$ in the outer halo in the bottom panel of $64 < r_{\rm e} < 128$ kpc, where the south one may be the local wake and the north one may be the collective halo response due to the dynamical reaction of the Galactic halo to the Large Magellanic Cloud \citep[e.g.][]{2019ApJ...884...51G, 2020MNRAS.498.5574E, 2021Natur.592..534C}.

Apart from the density profile, the composition of RRLs, particularly for the observed over-densities mentioned above, is interesting to study because it is likely related to their birth environment \citep{1973ApJ...185..477V, 1999AJ....118.1373L, 2004AJ....128..858S}. 
The period-amplitude diagram is typical to study the composition of RRLs and to verify the quality of a RRL catalogue, so we show the distribution of our RRLs in Figure~\ref{fig:cat_amp_period}, where the amplitude $A = \sqrt{A^2_r + A^2_g}$ with $A_r$ and $A_g$ defined in Equation~\ref{eq:amp_r_g}, and $P_{\rm best}$ is the best fitting period. 
We note that the location of a star in this diagram can be affected by the presence of the Blazhko effect \citep{1907AN....175..325B} or by the period aliasing during the Fourier fitting stage \citep{1976Ap&SS..39..447L, 1982ApJ...263..835S, 2018ApJS..236...16V}. 
There are two main clusters of the RRL type ab and c (RRab and RRc) roughly separated by the black dash-dotted line of $P_{\rm best} = 0.45$ days; the RRab cluster is to the right whereas the RRc cluster is to the left. 
It is worth noting that during the classification process, we never separate the two types of RRLs yet the classifier can still identify both of them. 
There are vertical patterns of RRLs at $P_{\rm best} = 0.33$ and 0.51 days, which are very likely caused by the aliasing period issue when fitting the light curves.
Also we note that the RRc stars might be contaminated by binary stars of the W Ursae Majoris type due to their sinusoidal light curves and period ranging between 0.25 and 0.6 days \citep{1998AJ....115.1135R}, which would be hard to distinguish with on our classification pipeline.

Looking closely at each cluster, we can see the Oosterhoff dichotomy \citep{1939Obs....62..104O, 2009Ap&SS.320..261C}, the more populated Oosterhoff I (OoI) and the less populated Oosterhoff II (OoII) that is shifted to longer periods given an amplitude. 
For the stars of the RRab cluster in our catalogue (whose periods are greater than 0.45 days), we compute the number counts on every grid of the period-amplitude plane, and utilize the grids with maximum number counts in each amplitude bin to fit a relation of $ A = -10.26 P_{\rm best}^2 + 8.27 P_{\rm best} - 0.88 $ to describe the distribution of OoI stars on the period-amplitude plane in Figure~\ref{fig:cat_amp_period}. 
Then we shift the curve by 0.025 days in the direction of period to roughly separate the OoI and OoII stars as 
\begin{equation} 
\label{eq:Oo_boundary}
    A = -10.26 \left( P_{\rm best} - 0.025 \right)^2 + 8.27 \left( P_{\rm best} - 0.025 \right) - 0.88
\end{equation}
which is shown by the black dashed curve in Figure~\ref{fig:cat_amp_period}. 
With the OoI RRLs to the left of the boundary and the OoII RRLs to the right of the boundary, we define the OoI fraction as $f_1 = N_1 / \left( N_1 + N_2 \right)$ where $N_1$ and $N_2$ are the numbers of OoI and OoII RRLs, finding the overall $f_1 = 0.65$.

According to the explained separation of OoI and OoII above, we are able to study the variation of OoI fraction $f_1$ across the Galactic halo, together with the RRL density distribution. 
The right panels of Figure~\ref{fig:cat_rrl_density_f1_re_slices} show $f_1$ on shells of different elliptical radii $r_{\rm e}$ in the coordinate of the Galactocentric longitude $\Phi$ and latitude $\Theta$ for the RRab stars in our catalogue. 
We note that for $64 < r_{\rm e} < 128$ kpc in the bottom panel, $f_1$ is so noisy that we grey it out.
Overall, $f_1$ is higher at the radii between $16 < r_{\rm e} < 64$ kpc, especially between $16 < r_{\rm e} < 32$ kpc which is roughly consistent with the finding in Figure 2 in \citet{2018MNRAS.477.1472B}. 
We observe that $f_1$ seems particularly anisotropic for $16 < r_{\rm e} < 32$ kpc. 
When looking at the locations of individual over-densities such as the Hercules-Aquila Cloud, the Virgo over-density and the Sagittarius Stream on the left panels, we observe no particular high or low $f_1$ corresponding to these over-densities in the right panels with the exception of somewhat higher  $f_1$  for the Hercules-Aquila Cloud in the $16 < r_{\rm e} < 32$ kpc distance range.

Another interesting point is the slightly higher $f_1$ around the solar neighbourhood $(\Phi, \Theta) = (180^\circ, 0^\circ)$ for $r_{\rm e} \sim 8$ kpc, which might be the Splash stars dubbed in \citet{2020MNRAS.494.3880B}, yet $f_1$ around the disk for $16 < r_{\rm e} < 32$ kpc is way higher than $f_1$ in the solar neighbourhood with $\Phi$ between $120^\circ$ to $210^\circ$ and $\Theta$ between $-30^\circ$ to $30^\circ$.

\section{Conclusions}
\label{sec:conclusion}

In this work, we have presented the RRL catalogue constructed from the combination of ZTF DR3 with \gaia EDR3, where \gaia provides accurate positions and proper motions on the whole sky and ZTF provides the vast amount of light curves with large epochs in multi-bands in the Northern sky. 
Starting from the source list in the join set of \gaia EDR3 and ZTF DR3 and the label of the SOS \gaia DR2 RRLs, we have processed them through the classification pipeline, that included the light curve fitting by a constant, single sinusoidal, third-order Fourier model in multiple bands, and two random forest classification steps to predict the probability for each source being a RRL.

Generating the RRL catalogue based on the predicted probability, we have obtained a catalogue that consists of 71,755 objects predicted to be RR Lyrae with at least 92\% purity and 92\% completeness compared to the SOS \gaia DR2 RRLs in the high galactic latitude areas with a high number of Gaia observations. 
The completeness of the RRL catalogue is generally higher than 80\% at the heliocentric distances closer than 80~kpc but drops drastically to 0 after 100 kpc. 
The catalogue covers the Northern sky above $-28^\circ$ in declination and the most distant RRL in it is at the heliocentric distance of 132~kpc. 
Compared with other RRL catalogues covering the Northern sky, the RRL catalogue of this work has more RRLs in the Galactic halo and is more complete at low Galactic latitude areas.

Using the new constructed RRL catalogue to analyze the Galactic halo density distribution, we observe the broadly ellipsoidal stellar distribution with flattening around 0.6 and power-law density profile with three known major over-densities of the halo substructure dominating: the Virgo over-density, the Hercules-Aquila Cloud, and the Sagittarius Stream. We do not observe a significant population associated with the Galactic disk \citep{Iorio2021}. The RRL density distribution seems to demonstrate the connection between the Virgo over-density and the Hercules-Aquila Cloud, supporting the possible association of several over-densities such as  Hercules-Aquila, Virgo, Eridanus–Phoenix and their link to the Gaia-Encelladus-Sausage merger \citep[i.e.][]{2019MNRAS.482..921S}. Besides, the RRL over-density in the Northern hemispheres is in broad agreement with the effect of the dynamical response of the Galactic halo to the Large Magellanic Cloud \citep[i.e.][]{2021Natur.592..534C}. We also analyse the Oosterhoff fraction differences across the halo, comparing it to the density distribution. We observe a higher fraction at the radii between $16 < r_{\rm e} < 32$ kpc with some anisotropy across the sky, but no clear association of this  with known major over-densities.

\section*{Acknowledgements}
SK was previously supported by NSF grants AST-1813881, AST-1909584, and Heising-Simons Foundation grant 2018-1030. This paper has made use of the Whole Sky Database (wsdb) created by Sergey Koposov and maintained at the Institute of Astronomy, Cambridge with financial support from the Science \& Technology Facilities Council (STFC) and the European Research Council (ERC). This paper has made use of the \code{q3c} software \citep{Koposov2006}.

This work presents results from the European Space Agency (ESA) space mission \gaia. \gaia data are being processed by the \gaia Data Processing and Analysis Consortium (DPAC). Funding for the DPAC is provided by national institutions, in particular the institutions participating in the \gaia MultiLateral Agreement (MLA). The \gaia mission website is \href{https://www.cosmos.esa.int/gaia}{https://www.cosmos.esa.int/gaia}. The \gaia archive website is \href{https://archives.esac.esa.int/gaia}{https://archives.esac.esa.int/gaia}.

Software:
\acknowledgeSoftware{python},
\acknowledgeSoftware{numpy},
\acknowledgeSoftware{scipy},
\acknowledgeSoftware{pandas},
\acknowledgeSoftware{matplotlib},
\acknowledgeSoftware{seaborn},
\acknowledgeSoftware{astropy},
\acknowledgeSoftware{sqlutilpy},
\code{healpy} \citep{2005ApJ...622..759G, Zonca2019}.

%%%%%%%%%%%%%%%%%%%%%%%%%%%%%%%%%%%%%%%%%%%%%%%%%%
\section*{Data Availability}

The data underlying this article were derived from sources in the public domain:
\begin{itemize}[leftmargin=*]
    \item ZTF DR3: \href{https://www.ztf.caltech.edu/page/dr3}{https://www.ztf.caltech.edu/page/dr3}
    \item \gaia EDR3: \href{https://archives.esac.esa.int/gaia}{https://archives.esac.esa.int/gaia}
    \item \gaia DR2 SOS RR Lyrae: \cite{2019A&A...622A..60C}
\end{itemize}



\bibliographystyle{mnras}
\bibliography{bib.tex} % if your bibtex file is called example.bib

\begin{thebibliography}{}
\makeatletter
\relax
\def\mn@urlcharsother{\let\do\@makeother \do\$\do\&\do\#\do\^\do\_\do\%\do\~}
\def\mn@doi{\begingroup\mn@urlcharsother \@ifnextchar [ {\mn@doi@}
  {\mn@doi@[]}}
\def\mn@doi@[#1]#2{\def\@tempa{#1}\ifx\@tempa\@empty \href
  {http://dx.doi.org/#2} {doi:#2}\else \href {http://dx.doi.org/#2} {#1}\fi
  \endgroup}
\def\mn@eprint#1#2{\mn@eprint@#1:#2::\@nil}
\def\mn@eprint@arXiv#1{\href {http://arxiv.org/abs/#1} {{\tt arXiv:#1}}}
\def\mn@eprint@dblp#1{\href {http://dblp.uni-trier.de/rec/bibtex/#1.xml}
  {dblp:#1}}
\def\mn@eprint@#1:#2:#3:#4\@nil{\def\@tempa {#1}\def\@tempb {#2}\def\@tempc
  {#3}\ifx \@tempc \@empty \let \@tempc \@tempb \let \@tempb \@tempa \fi \ifx
  \@tempb \@empty \def\@tempb {arXiv}\fi \@ifundefined
  {mn@eprint@\@tempb}{\@tempb:\@tempc}{\expandafter \expandafter \csname
  mn@eprint@\@tempb\endcsname \expandafter{\@tempc}}}

\bibitem[\protect\citeauthoryear{{Astropy Collaboration} et~al.,}{{Astropy
  Collaboration} et~al.}{2013}]{astropy}
{Astropy Collaboration} et~al., 2013, \mn@doi [\aap]
  {10.1051/0004-6361/201322068}, \href
  {http://adsabs.harvard.edu/abs/2013A%26A...558A..33A} {558, A33}

\bibitem[\protect\citeauthoryear{Baker \& Willman}{Baker \&
  Willman}{2015}]{Baker_2015}
Baker M.,  Willman B.,  2015, \mn@doi [The Astronomical Journal]
  {10.1088/0004-6256/150/5/160}, 150, 160

\bibitem[\protect\citeauthoryear{{Balbinot} \& {Helmi}}{{Balbinot} \&
  {Helmi}}{2021}]{2021arXiv210409794B}
{Balbinot} E.,  {Helmi} A.,  2021, arXiv e-prints, \href
  {https://ui.adsabs.harvard.edu/abs/2021arXiv210409794B} {p. arXiv:2104.09794}

\bibitem[\protect\citeauthoryear{{Bellm} et~al.,}{{Bellm}
  et~al.}{2019}]{2019PASP..131a8002B}
{Bellm} E.~C.,  et~al., 2019, \mn@doi [\pasp] {10.1088/1538-3873/aaecbe}, \href
  {https://ui.adsabs.harvard.edu/abs/2019PASP..131a8002B} {131, 018002}

\bibitem[\protect\citeauthoryear{{Belokurov} et~al.,}{{Belokurov}
  et~al.}{2007}]{2007ApJ...657L..89B}
{Belokurov} V.,  et~al., 2007, \mn@doi [\apjl] {10.1086/513144}, \href
  {https://ui.adsabs.harvard.edu/abs/2007ApJ...657L..89B} {657, L89}

\bibitem[\protect\citeauthoryear{{Belokurov}, {Deason}, {Koposov}, {Catelan},
  {Erkal}, {Drake}  \& {Evans}}{{Belokurov} et~al.}{2018}]{2018MNRAS.477.1472B}
{Belokurov} V.,  {Deason} A.~J.,  {Koposov} S.~E.,  {Catelan} M.,  {Erkal} D.,
  {Drake} A.~J.,   {Evans} N.~W.,  2018, \mn@doi [\mnras]
  {10.1093/mnras/sty615}, \href
  {https://ui.adsabs.harvard.edu/abs/2018MNRAS.477.1472B} {477, 1472}

\bibitem[\protect\citeauthoryear{{Belokurov}, {Sanders}, {Fattahi}, {Smith},
  {Deason}, {Evans}  \& {Grand}}{{Belokurov}
  et~al.}{2020}]{2020MNRAS.494.3880B}
{Belokurov} V.,  {Sanders} J.~L.,  {Fattahi} A.,  {Smith} M.~C.,  {Deason}
  A.~J.,  {Evans} N.~W.,   {Grand} R. J.~J.,  2020, \mn@doi [\mnras]
  {10.1093/mnras/staa876}, \href
  {https://ui.adsabs.harvard.edu/abs/2020MNRAS.494.3880B} {494, 3880}

\bibitem[\protect\citeauthoryear{{Bla{\v{z}}ko}}{{Bla{\v{z}}ko}}{1907}]{1907AN....175..325B}
{Bla{\v{z}}ko} S.,  1907, \mn@doi [Astronomische Nachrichten]
  {10.1002/asna.19071752002}, \href
  {https://ui.adsabs.harvard.edu/abs/1907AN....175..325B} {175, 325}

\bibitem[\protect\citeauthoryear{{Bonaca} et~al.,}{{Bonaca}
  et~al.}{2012}]{2012AJ....143..105B}
{Bonaca} A.,  et~al., 2012, \mn@doi [\aj] {10.1088/0004-6256/143/5/105}, \href
  {https://ui.adsabs.harvard.edu/abs/2012AJ....143..105B} {143, 105}

\bibitem[\protect\citeauthoryear{Breiman}{Breiman}{2001}]{Statistics01randomforests}
Breiman L.,  2001, in Machine Learning. pp 5--32

\bibitem[\protect\citeauthoryear{Cacciari \& Clementini}{Cacciari \&
  Clementini}{2003}]{Cacciari2003}
Cacciari C.,  Clementini G.,  2003, Globular Cluster Distances from RR Lyrae
  Stars.
Springer Berlin Heidelberg, Berlin, Heidelberg, pp 105--122,
  \mn@doi{10.1007/978-3-540-39882-0_6}, \url
  {https://doi.org/10.1007/978-3-540-39882-0_6}

\bibitem[\protect\citeauthoryear{C{\'{a}}ceres \& Catelan}{C{\'{a}}ceres \&
  Catelan}{2008}]{C_ceres_2008}
C{\'{a}}ceres C.,  Catelan M.,  2008, \mn@doi [The Astrophysical Journal
  Supplement Series] {10.1086/591231}, 179, 242

\bibitem[\protect\citeauthoryear{{Catelan}}{{Catelan}}{2009}]{2009Ap&SS.320..261C}
{Catelan} M.,  2009, \mn@doi [\apss] {10.1007/s10509-009-9987-8}, \href
  {https://ui.adsabs.harvard.edu/abs/2009Ap&SS.320..261C} {320, 261}

\bibitem[\protect\citeauthoryear{Catelan, Pritzl  \& Smith}{Catelan
  et~al.}{2004}]{Catelan_2004}
Catelan M.,  Pritzl B.~J.,   Smith H.~A.,  2004, \mn@doi [The Astrophysical
  Journal Supplement Series] {10.1086/422916}, 154, 633

\bibitem[\protect\citeauthoryear{{Chen}, {Wang}, {Deng}, {de Grijs}, {Yang}  \&
  {Tian}}{{Chen} et~al.}{2020}]{2020ApJS..249...18C}
{Chen} X.,  {Wang} S.,  {Deng} L.,  {de Grijs} R.,  {Yang} M.,   {Tian} H.,
  2020, \mn@doi [\apjs] {10.3847/1538-4365/ab9cae}, \href
  {https://ui.adsabs.harvard.edu/abs/2020ApJS..249...18C} {249, 18}

\bibitem[\protect\citeauthoryear{{Clementini} et~al.,}{{Clementini}
  et~al.}{2019a}]{2019A&A...622A..60C}
{Clementini} G.,  et~al., 2019a, \mn@doi [\aap] {10.1051/0004-6361/201833374},
  \href {https://ui.adsabs.harvard.edu/abs/2019A&A...622A..60C} {622, A60}

\bibitem[\protect\citeauthoryear{{Clementini} et~al.,}{{Clementini}
  et~al.}{2019b}]{2019AA...622A..60C}
{Clementini} G.,  et~al., 2019b, \mn@doi [\aap] {10.1051/0004-6361/201833374},
  \href {https://ui.adsabs.harvard.edu/abs/2019A&A...622A..60C} {622, A60}

\bibitem[\protect\citeauthoryear{{Conroy}, {Naidu}, {Garavito-Camargo},
  {Besla}, {Zaritsky}, {Bonaca}  \& {Johnson}}{{Conroy}
  et~al.}{2021}]{2021Natur.592..534C}
{Conroy} C.,  {Naidu} R.~P.,  {Garavito-Camargo} N.,  {Besla} G.,  {Zaritsky}
  D.,  {Bonaca} A.,   {Johnson} B.~D.,  2021, \mn@doi [\nat]
  {10.1038/s41586-021-03385-7}, \href
  {https://ui.adsabs.harvard.edu/abs/2021Natur.592..534C} {592, 534}

\bibitem[\protect\citeauthoryear{{Drake} et~al.,}{{Drake}
  et~al.}{2009}]{2009ApJ...696..870D}
{Drake} A.~J.,  et~al., 2009, \mn@doi [\apj] {10.1088/0004-637X/696/1/870},
  \href {https://ui.adsabs.harvard.edu/abs/2009ApJ...696..870D} {696, 870}

\bibitem[\protect\citeauthoryear{{Drake} et~al.,}{{Drake}
  et~al.}{2014}]{2014ApJS..213....9D}
{Drake} A.~J.,  et~al., 2014, \mn@doi [\apjs] {10.1088/0067-0049/213/1/9},
  \href {https://ui.adsabs.harvard.edu/abs/2014ApJS..213....9D} {213, 9}

\bibitem[\protect\citeauthoryear{{Duffau}, {Zinn}, {Vivas}, {Carraro},
  {M{\'e}ndez}, {Winnick}  \& {Gallart}}{{Duffau}
  et~al.}{2006}]{2006ApJ...636L..97D}
{Duffau} S.,  {Zinn} R.,  {Vivas} A.~K.,  {Carraro} G.,  {M{\'e}ndez} R.~A.,
  {Winnick} R.,   {Gallart} C.,  2006, \mn@doi [\apjl] {10.1086/500130}, \href
  {https://ui.adsabs.harvard.edu/abs/2006ApJ...636L..97D} {636, L97}

\bibitem[\protect\citeauthoryear{{Erkal}, {Belokurov}  \& {Parkin}}{{Erkal}
  et~al.}{2020}]{2020MNRAS.498.5574E}
{Erkal} D.,  {Belokurov} V.~A.,   {Parkin} D.~L.,  2020, \mn@doi [\mnras]
  {10.1093/mnras/staa2840}, \href
  {https://ui.adsabs.harvard.edu/abs/2020MNRAS.498.5574E} {498, 5574}

\bibitem[\protect\citeauthoryear{{Fiorentino} et~al.,}{{Fiorentino}
  et~al.}{2015}]{2015ApJ...798L..12F}
{Fiorentino} G.,  et~al., 2015, \mn@doi [\apjl] {10.1088/2041-8205/798/1/L12},
  \href {https://ui.adsabs.harvard.edu/abs/2015ApJ...798L..12F} {798, L12}

\bibitem[\protect\citeauthoryear{{Gaia Collaboration} et~al.,}{{Gaia
  Collaboration} et~al.}{2016}]{2016A&A...595A...1G}
{Gaia Collaboration} et~al., 2016, \mn@doi [\aap]
  {10.1051/0004-6361/201629272}, \href
  {https://ui.adsabs.harvard.edu/abs/2016A&A...595A...1G} {595, A1}

\bibitem[\protect\citeauthoryear{{Gaia Collaboration}, {Brown}, {Vallenari},
  {Prusti}, {de Bruijne}, {Babusiaux}  \& {Biermann}}{{Gaia Collaboration}
  et~al.}{2020}]{2020arXiv201201533G}
{Gaia Collaboration} {Brown} A.~G.~A.,  {Vallenari} A.,  {Prusti} T.,  {de
  Bruijne} J.~H.~J.,  {Babusiaux} C.,   {Biermann} M.,  2020, arXiv e-prints,
  \href {https://ui.adsabs.harvard.edu/abs/2020arXiv201201533G} {p.
  arXiv:2012.01533}

\bibitem[\protect\citeauthoryear{{Garavito-Camargo}, {Besla}, {Laporte},
  {Johnston}, {G{\'o}mez}  \& {Watkins}}{{Garavito-Camargo}
  et~al.}{2019}]{2019ApJ...884...51G}
{Garavito-Camargo} N.,  {Besla} G.,  {Laporte} C. F.~P.,  {Johnston} K.~V.,
  {G{\'o}mez} F.~A.,   {Watkins} L.~L.,  2019, \mn@doi [\apj]
  {10.3847/1538-4357/ab32eb}, \href
  {https://ui.adsabs.harvard.edu/abs/2019ApJ...884...51G} {884, 51}

\bibitem[\protect\citeauthoryear{{G{\'o}rski}, {Hivon}, {Banday}, {Wandelt},
  {Hansen}, {Reinecke}  \& {Bartelmann}}{{G{\'o}rski}
  et~al.}{2005}]{2005ApJ...622..759G}
{G{\'o}rski} K.~M.,  {Hivon} E.,  {Banday} A.~J.,  {Wandelt} B.~D.,  {Hansen}
  F.~K.,  {Reinecke} M.,   {Bartelmann} M.,  2005, \mn@doi [\apj]
  {10.1086/427976}, \href {http://adsabs.harvard.edu/abs/2005ApJ...622..759G}
  {622, 759}

\bibitem[\protect\citeauthoryear{{Harris}}{{Harris}}{1996}]{1996AJ....112.1487H}
{Harris} W.~E.,  1996, \mn@doi [\aj] {10.1086/118116}, \href
  {https://ui.adsabs.harvard.edu/abs/1996AJ....112.1487H} {112, 1487}

\bibitem[\protect\citeauthoryear{{Harris}}{{Harris}}{2010}]{2010arXiv1012.3224H}
{Harris} W.~E.,  2010, arXiv e-prints, \href
  {https://ui.adsabs.harvard.edu/abs/2010arXiv1012.3224H} {p. arXiv:1012.3224}

\bibitem[\protect\citeauthoryear{{Hernitschek} et~al.,}{{Hernitschek}
  et~al.}{2016}]{2016ApJ...817...73H}
{Hernitschek} N.,  et~al., 2016, \mn@doi [\apj] {10.3847/0004-637X/817/1/73},
  \href {https://ui.adsabs.harvard.edu/abs/2016ApJ...817...73H} {817, 73}

\bibitem[\protect\citeauthoryear{{Hernitschek} et~al.,}{{Hernitschek}
  et~al.}{2017}]{2017ApJ...850...96H}
{Hernitschek} N.,  et~al., 2017, \mn@doi [\apj] {10.3847/1538-4357/aa960c},
  \href {https://ui.adsabs.harvard.edu/abs/2017ApJ...850...96H} {850, 96}

\bibitem[\protect\citeauthoryear{{Holl} et~al.,}{{Holl}
  et~al.}{2018}]{2018A&A...618A..30H}
{Holl} B.,  et~al., 2018, \mn@doi [\aap] {10.1051/0004-6361/201832892}, \href
  {https://ui.adsabs.harvard.edu/abs/2018A&A...618A..30H} {618, A30}

\bibitem[\protect\citeauthoryear{Homma et~al.,}{Homma
  et~al.}{2016}]{Homma_2016}
Homma D.,  et~al., 2016, \mn@doi [The Astrophysical Journal]
  {10.3847/0004-637x/832/1/21}, 832, 21

\bibitem[\protect\citeauthoryear{Huang \& Koposov}{Huang \&
  Koposov}{2021}]{huang_kuan_wei_2021_5774018}
Huang K.-W.,  Koposov S.~E.,  2021, The RR Lyrae variable catalog of ZTF DR3,
  \mn@doi{10.5281/zenodo.5774018}, \url
  {https://doi.org/10.5281/zenodo.5774018}

\bibitem[\protect\citeauthoryear{Hunter}{Hunter}{2007}]{matplotlib}
Hunter J.~D.,  2007, \mn@doi [Computing in Science \& Engineering]
  {http://dx.doi.org/10.1109/MCSE.2007.55}, 9, 90

\bibitem[\protect\citeauthoryear{{Iorio} \& {Belokurov}}{{Iorio} \&
  {Belokurov}}{2021}]{Iorio2021}
{Iorio} G.,  {Belokurov} V.,  2021, \mn@doi [\mnras] {10.1093/mnras/stab005},
  \href {https://ui.adsabs.harvard.edu/abs/2021MNRAS.502.5686I} {502, 5686}

\bibitem[\protect\citeauthoryear{{Iorio}, {Belokurov}, {Erkal}, {Koposov},
  {Nipoti}  \& {Fraternali}}{{Iorio} et~al.}{2018}]{2018MNRAS.474.2142I}
{Iorio} G.,  {Belokurov} V.,  {Erkal} D.,  {Koposov} S.~E.,  {Nipoti} C.,
  {Fraternali} F.,  2018, \mn@doi [\mnras] {10.1093/mnras/stx2819}, \href
  {https://ui.adsabs.harvard.edu/abs/2018MNRAS.474.2142I} {474, 2142}

\bibitem[\protect\citeauthoryear{{Ivezi{\'c}} et~al.,}{{Ivezi{\'c}}
  et~al.}{2008}]{2008ApJ...684..287I}
{Ivezi{\'c}} {\v{Z}}.,  et~al., 2008, \mn@doi [\apj] {10.1086/589678}, \href
  {https://ui.adsabs.harvard.edu/abs/2008ApJ...684..287I} {684, 287}

\bibitem[\protect\citeauthoryear{{Jayasinghe} et~al.,}{{Jayasinghe}
  et~al.}{2018}]{2018RNAAS...2...18J}
{Jayasinghe} T.,  et~al., 2018, \mn@doi [Research Notes of the American
  Astronomical Society] {10.3847/2515-5172/aaaa20}, \href
  {https://ui.adsabs.harvard.edu/abs/2018RNAAS...2...18J} {2, 18}

\bibitem[\protect\citeauthoryear{{Jayasinghe} et~al.,}{{Jayasinghe}
  et~al.}{2020}]{2020yCat.2366....0J}
{Jayasinghe} T.,  et~al., 2020, VizieR Online Data Catalog, \href
  {https://ui.adsabs.harvard.edu/abs/2020yCat.2366....0J} {p. II/366}

\bibitem[\protect\citeauthoryear{Jones, Oliphant, Peterson  et~al.}{Jones
  et~al.}{2001}]{scipy}
Jones E.,  Oliphant T.,  Peterson P.,   et~al., 2001, {SciPy}: Open source
  scientific tools for {Python}, \url {http://www.scipy.org/}

\bibitem[\protect\citeauthoryear{{Jurcsik} \& {Kovacs}}{{Jurcsik} \&
  {Kovacs}}{1996}]{1996A&A...312..111J}
{Jurcsik} J.,  {Kovacs} G.,  1996, \aap, \href
  {https://ui.adsabs.harvard.edu/abs/1996A&A...312..111J} {312, 111}

\bibitem[\protect\citeauthoryear{{Juri{\'c}} et~al.,}{{Juri{\'c}}
  et~al.}{2008}]{2008ApJ...673..864J}
{Juri{\'c}} M.,  et~al., 2008, \mn@doi [\apj] {10.1086/523619}, \href
  {https://ui.adsabs.harvard.edu/abs/2008ApJ...673..864J} {673, 864}

\bibitem[\protect\citeauthoryear{Koposov}{Koposov}{2021}]{sqlutilpy}
Koposov S.,  2021, segasai/sqlutilpy: sqlutilpy v0.16.0,
  \mn@doi{10.5281/zenodo.5160119}, \url
  {https://doi.org/10.5281/zenodo.5160119}

\bibitem[\protect\citeauthoryear{{Koposov} \& {Bartunov}}{{Koposov} \&
  {Bartunov}}{2006}]{Koposov2006}
{Koposov} S.,  {Bartunov} O.,  2006, in {Gabriel} C.,  {Arviset} C.,  {Ponz}
  D.,   {Enrique} S.,  eds,  Astronomical Society of the Pacific Conference
  Series Vol. 351, Astronomical Data Analysis Software and Systems XV. p.~735

\bibitem[\protect\citeauthoryear{{Lee} \& {Carney}}{{Lee} \&
  {Carney}}{1999}]{1999AJ....118.1373L}
{Lee} J.-W.,  {Carney} B.~W.,  1999, \mn@doi [\aj] {10.1086/301008}, \href
  {https://ui.adsabs.harvard.edu/abs/1999AJ....118.1373L} {118, 1373}

\bibitem[\protect\citeauthoryear{{Li} et~al.,}{{Li}
  et~al.}{2016}]{2016ApJ...817..135L}
{Li} T.~S.,  et~al., 2016, \mn@doi [\apj] {10.3847/0004-637X/817/2/135}, \href
  {https://ui.adsabs.harvard.edu/abs/2016ApJ...817..135L} {817, 135}

\bibitem[\protect\citeauthoryear{{Lomb}}{{Lomb}}{1976}]{1976Ap&SS..39..447L}
{Lomb} N.~R.,  1976, \mn@doi [\apss] {10.1007/BF00648343}, \href
  {https://ui.adsabs.harvard.edu/abs/1976Ap&SS..39..447L} {39, 447}

\bibitem[\protect\citeauthoryear{{Marconi}}{{Marconi}}{2012}]{2012MSAIS..19..138M}
{Marconi} M.,  2012, Memorie della Societa Astronomica Italiana Supplementi,
  \href {https://ui.adsabs.harvard.edu/abs/2012MSAIS..19..138M} {19, 138}

\bibitem[\protect\citeauthoryear{Martínez-Vázquez et~al.,}{Martínez-Vázquez
  et~al.}{2019}]{10.1093/mnras/stz2609}
Martínez-Vázquez C.~E.,  et~al., 2019, \mn@doi [Monthly Notices of the Royal
  Astronomical Society] {10.1093/mnras/stz2609}, 490, 2183

\bibitem[\protect\citeauthoryear{{Masci} et~al.,}{{Masci}
  et~al.}{2019}]{2019PASP..131a8003M}
{Masci} F.~J.,  et~al., 2019, \mn@doi [\pasp] {10.1088/1538-3873/aae8ac}, \href
  {https://ui.adsabs.harvard.edu/abs/2019PASP..131a8003M} {131, 018003}

\bibitem[\protect\citeauthoryear{{Massari} \& {Helmi}}{{Massari} \&
  {Helmi}}{2018}]{2018AA...620A.155M}
{Massari} D.,  {Helmi} A.,  2018, \mn@doi [\aap] {10.1051/0004-6361/201833367},
  \href {https://ui.adsabs.harvard.edu/abs/2018A&A...620A.155M} {620, A155}

\bibitem[\protect\citeauthoryear{{McConnachie}}{{McConnachie}}{2012}]{2012AJ....144....4M}
{McConnachie} A.~W.,  2012, \mn@doi [\aj] {10.1088/0004-6256/144/1/4}, \href
  {https://ui.adsabs.harvard.edu/abs/2012AJ....144....4M} {144, 4}

\bibitem[\protect\citeauthoryear{Mckinney}{Mckinney}{2010}]{pandas}
Mckinney W.,  2010, Data Structures for Statistical Computing in Python,
  \mn@doi{10.25080/Majora-92bf1922-00a}

\bibitem[\protect\citeauthoryear{{Medina} et~al.,}{{Medina}
  et~al.}{2018}]{2018ApJ...855...43M}
{Medina} G.~E.,  et~al., 2018, \mn@doi [\apj] {10.3847/1538-4357/aaad02}, \href
  {https://ui.adsabs.harvard.edu/abs/2018ApJ...855...43M} {855, 43}

\bibitem[\protect\citeauthoryear{{Newberg} et~al.,}{{Newberg}
  et~al.}{2002}]{2002ApJ...569..245N}
{Newberg} H.~J.,  et~al., 2002, \mn@doi [\apj] {10.1086/338983}, \href
  {https://ui.adsabs.harvard.edu/abs/2002ApJ...569..245N} {569, 245}

\bibitem[\protect\citeauthoryear{{Oosterhoff}}{{Oosterhoff}}{1939}]{1939Obs....62..104O}
{Oosterhoff} P.~T.,  1939, The Observatory, \href
  {https://ui.adsabs.harvard.edu/abs/1939Obs....62..104O} {62, 104}

\bibitem[\protect\citeauthoryear{Pedregosa et~al.,}{Pedregosa
  et~al.}{2011}]{scikit-learn}
Pedregosa F.,  et~al., 2011, Journal of Machine Learning Research, 12, 2825

\bibitem[\protect\citeauthoryear{{Rucinski}}{{Rucinski}}{1998}]{1998AJ....115.1135R}
{Rucinski} S.~M.,  1998, \mn@doi [\aj] {10.1086/300266}, \href
  {https://ui.adsabs.harvard.edu/abs/1998AJ....115.1135R} {115, 1135}

\bibitem[\protect\citeauthoryear{{Sandage}}{{Sandage}}{2004}]{2004AJ....128..858S}
{Sandage} A.,  2004, \mn@doi [\aj] {10.1086/422509}, \href
  {https://ui.adsabs.harvard.edu/abs/2004AJ....128..858S} {128, 858}

\bibitem[\protect\citeauthoryear{{Scargle}}{{Scargle}}{1982}]{1982ApJ...263..835S}
{Scargle} J.~D.,  1982, \mn@doi [\apj] {10.1086/160554}, \href
  {https://ui.adsabs.harvard.edu/abs/1982ApJ...263..835S} {263, 835}

\bibitem[\protect\citeauthoryear{{Schlafly} \& {Finkbeiner}}{{Schlafly} \&
  {Finkbeiner}}{2011}]{2011ApJ...737..103S}
{Schlafly} E.~F.,  {Finkbeiner} D.~P.,  2011, \mn@doi [\apj]
  {10.1088/0004-637X/737/2/103}, \href
  {https://ui.adsabs.harvard.edu/abs/2011ApJ...737..103S} {737, 103}

\bibitem[\protect\citeauthoryear{{Sesar} et~al.,}{{Sesar}
  et~al.}{2010}]{2010ApJ...708..717S}
{Sesar} B.,  et~al., 2010, \mn@doi [\apj] {10.1088/0004-637X/708/1/717}, \href
  {https://ui.adsabs.harvard.edu/abs/2010ApJ...708..717S} {708, 717}

\bibitem[\protect\citeauthoryear{Sesar et~al.,}{Sesar
  et~al.}{2014}]{Sesar_2014}
Sesar B.,  et~al., 2014, \mn@doi [The Astrophysical Journal]
  {10.1088/0004-637x/793/2/135}, 793, 135

\bibitem[\protect\citeauthoryear{{Sesar} et~al.,}{{Sesar}
  et~al.}{2017}]{2017AJ....153..204S}
{Sesar} B.,  et~al., 2017, \mn@doi [\aj] {10.3847/1538-3881/aa661b}, \href
  {https://ui.adsabs.harvard.edu/abs/2017AJ....153..204S} {153, 204}

\bibitem[\protect\citeauthoryear{{Simion}, {Belokurov}, {Irwin}  \&
  {Koposov}}{{Simion} et~al.}{2014}]{2014MNRAS.440..161S}
{Simion} I.~T.,  {Belokurov} V.,  {Irwin} M.,   {Koposov} S.~E.,  2014, \mn@doi
  [\mnras] {10.1093/mnras/stu133}, \href
  {https://ui.adsabs.harvard.edu/abs/2014MNRAS.440..161S} {440, 161}

\bibitem[\protect\citeauthoryear{{Simion}, {Belokurov}, {Koposov}, {Sheffield}
  \& {Johnston}}{{Simion} et~al.}{2018}]{2018MNRAS.476.3913S}
{Simion} I.~T.,  {Belokurov} V.,  {Koposov} S.~E.,  {Sheffield} A.,
  {Johnston} K.~V.,  2018, \mn@doi [\mnras] {10.1093/mnras/sty499}, \href
  {https://ui.adsabs.harvard.edu/abs/2018MNRAS.476.3913S} {476, 3913}

\bibitem[\protect\citeauthoryear{{Simion}, {Belokurov}  \& {Koposov}}{{Simion}
  et~al.}{2019}]{2019MNRAS.482..921S}
{Simion} I.~T.,  {Belokurov} V.,   {Koposov} S.~E.,  2019, \mn@doi [\mnras]
  {10.1093/mnras/sty2744}, \href
  {https://ui.adsabs.harvard.edu/abs/2019MNRAS.482..921S} {482, 921}

\bibitem[\protect\citeauthoryear{{Simon} \& {Clement}}{{Simon} \&
  {Clement}}{1993}]{1993ApJ...410..526S}
{Simon} N.~R.,  {Clement} C.~M.,  1993, \mn@doi [\apj] {10.1086/172771}, \href
  {https://ui.adsabs.harvard.edu/abs/1993ApJ...410..526S} {410, 526}

\bibitem[\protect\citeauthoryear{{Smith}}{{Smith}}{1995}]{1995CAS....27.....S}
{Smith} H.~A.,  1995, Cambridge Astrophysics Series, \href
  {https://ui.adsabs.harvard.edu/abs/1995CAS....27.....S} {27}

\bibitem[\protect\citeauthoryear{{Soszy{\'n}ski} et~al.,}{{Soszy{\'n}ski}
  et~al.}{2019}]{2019AcA....69..321S}
{Soszy{\'n}ski} I.,  et~al., 2019, \mn@doi [\actaa]
  {10.32023/0001-5237/69.4.2}, \href
  {https://ui.adsabs.harvard.edu/abs/2019AcA....69..321S} {69, 321}

\bibitem[\protect\citeauthoryear{{Stetson}, {Fiorentino}, {Bono}, {Bernard},
  {Monelli}, {Iannicola}, {Gallart}  \& {Ferraro}}{{Stetson}
  et~al.}{2014}]{2014PASP..126..616S}
{Stetson} P.~B.,  {Fiorentino} G.,  {Bono} G.,  {Bernard} E.~J.,  {Monelli} M.,
   {Iannicola} G.,  {Gallart} C.,   {Ferraro} I.,  2014, \mn@doi [\pasp]
  {10.1086/677352}, \href
  {https://ui.adsabs.harvard.edu/abs/2014PASP..126..616S} {126, 616}

\bibitem[\protect\citeauthoryear{{Stringer} et~al.,}{{Stringer}
  et~al.}{2021}]{2021ApJ...911..109S}
{Stringer} K.~M.,  et~al., 2021, \mn@doi [\apj] {10.3847/1538-4357/abe873},
  \href {https://ui.adsabs.harvard.edu/abs/2021ApJ...911..109S} {911, 109}

\bibitem[\protect\citeauthoryear{{Torrealba} et~al.,}{{Torrealba}
  et~al.}{2015}]{2015MNRAS.446.2251T}
{Torrealba} G.,  et~al., 2015, \mn@doi [\mnras] {10.1093/mnras/stu2274}, \href
  {https://ui.adsabs.harvard.edu/abs/2015MNRAS.446.2251T} {446, 2251}

\bibitem[\protect\citeauthoryear{{Torrealba} et~al.,}{{Torrealba}
  et~al.}{2019}]{2019MNRAS.488.2743T}
{Torrealba} G.,  et~al., 2019, \mn@doi [\mnras] {10.1093/mnras/stz1624}, \href
  {https://ui.adsabs.harvard.edu/abs/2019MNRAS.488.2743T} {488, 2743}

\bibitem[\protect\citeauthoryear{Van~Rossum \& Drake}{Van~Rossum \&
  Drake}{2009}]{python}
Van~Rossum G.,  Drake F.~L.,  2009, Python 3 Reference Manual.
CreateSpace, Scotts Valley, CA

\bibitem[\protect\citeauthoryear{{VanderPlas}}{{VanderPlas}}{2018}]{2018ApJS..236...16V}
{VanderPlas} J.~T.,  2018, \mn@doi [\apjs] {10.3847/1538-4365/aab766}, \href
  {https://ui.adsabs.harvard.edu/abs/2018ApJS..236...16V} {236, 16}

\bibitem[\protect\citeauthoryear{Vivas \& Zinn}{Vivas \&
  Zinn}{2006}]{Vivas_2006}
Vivas A.~K.,  Zinn R.,  2006, \mn@doi [The Astronomical Journal]
  {10.1086/505200}, 132, 714

\bibitem[\protect\citeauthoryear{{Vivas} et~al.,}{{Vivas}
  et~al.}{2001}]{2001ApJ...554L..33V}
{Vivas} A.~K.,  et~al., 2001, \mn@doi [\apjl] {10.1086/320915}, \href
  {https://ui.adsabs.harvard.edu/abs/2001ApJ...554L..33V} {554, L33}

\bibitem[\protect\citeauthoryear{{Vivas} et~al.,}{{Vivas}
  et~al.}{2004}]{2004AJ....127.1158V}
{Vivas} A.~K.,  et~al., 2004, \mn@doi [\aj] {10.1086/380929}, \href
  {https://ui.adsabs.harvard.edu/abs/2004AJ....127.1158V} {127, 1158}

\bibitem[\protect\citeauthoryear{Waskom et~al.,}{Waskom et~al.}{2016}]{seaborn}
Waskom M.,  et~al., 2016, seaborn: v0.7.0 (January 2016),
  \mn@doi{10.5281/zenodo.45133}, \url {http://dx.doi.org/10.5281/zenodo.45133}

\bibitem[\protect\citeauthoryear{{Wils}, {Lloyd}  \& {Bernhard}}{{Wils}
  et~al.}{2006}]{2006MNRAS.368.1757W}
{Wils} P.,  {Lloyd} C.,   {Bernhard} K.,  2006, \mn@doi [\mnras]
  {10.1111/j.1365-2966.2006.10236.x}, \href
  {https://ui.adsabs.harvard.edu/abs/2006MNRAS.368.1757W} {368, 1757}

\bibitem[\protect\citeauthoryear{Zonca, Singer, Lenz, Reinecke, Rosset, Hivon
  \& Gorski}{Zonca et~al.}{2019}]{Zonca2019}
Zonca A.,  Singer L.,  Lenz D.,  Reinecke M.,  Rosset C.,  Hivon E.,   Gorski
  K.,  2019, \mn@doi [Journal of Open Source Software] {10.21105/joss.01298},
  4, 1298

\bibitem[\protect\citeauthoryear{{van Albada} \& {Baker}}{{van Albada} \&
  {Baker}}{1973}]{1973ApJ...185..477V}
{van Albada} T.~S.,  {Baker} N.,  1973, \mn@doi [\apj] {10.1086/152434}, \href
  {https://ui.adsabs.harvard.edu/abs/1973ApJ...185..477V} {185, 477}

\bibitem[\protect\citeauthoryear{{van~der~Walt}, Colbert  \&
  Varoquaux}{{van~der~Walt} et~al.}{2011}]{numpy}
{van~der~Walt} S.,  Colbert S.~C.,   Varoquaux G.,  2011, \mn@doi [Computing in
  Science \& Engineering] {http://dx.doi.org/10.1109/MCSE.2011.37}, 13, 22

\makeatother
\end{thebibliography}

%\begin{thebibliography}{99}
%\end{thebibliography}

%%%%%%%%%%%%%%%%%%%%%%%%%%%%%%%%%%%%%%%%%%%%%%%%%%

%%%%%%%%%%%%%%%%% APPENDICES %%%%%%%%%%%%%%%%%%%%%

% \appendix

% \section{Some extra material}

% If you want to present additional material which would interrupt the flow of the main paper,
% it can be placed in an Appendix which appears after the list of references.

%%%%%%%%%%%%%%%%%%%%%%%%%%%%%%%%%%%%%%%%%%%%%%%%%%

% Don't change these lines
\bsp	% typesetting comment
\label{lastpage}
\end{document}